\documentclass[%
 reprint,
superscriptaddress,
 amsmath,amssymb,
 aps,
]{revtex4-2}

\usepackage{graphicx}
\usepackage{dcolumn}
\usepackage{bm}

\usepackage[dvipsnames]{xcolor}
\usepackage{bm}
\usepackage{comment}
\usepackage{chemformula}

\begin{document}

\title{Loops are Geometric Catalysts for DNA Integration}

\author{Cleis Battaglia}
\affiliation{School of Physics and Astronomy, University of Edinburgh, Peter Guthrie Tait Road, Edinburgh, EH9 3FD, UK}
\author{Davide Michieletto}
\thanks{corresponding author, davide.michieletto@ed.ac.uk}
\affiliation{School of Physics and Astronomy, University of Edinburgh, Peter Guthrie Tait Road, Edinburgh, EH9 3FD, UK}
\affiliation{MRC Human Genetics Unit, Institute of Genetics and Cancer, University of Edinburgh, Edinburgh EH4 2XU, UK}

\begin{abstract}
\textbf{
The insertion of HIV and other DNA elements within genomes underpins both genetic diversity and disease when unregulated. Most of these insertions are not random and occupy specific positions within the genome but the physical mechanisms underlying the integration site selection are poorly understood. Here we perform Molecular Dynamics simulations to study the insertion of DNA elements, such as HIV viral DNA or transposons, into naked DNA or chromatin substrate. More specifically, we explore the role of loops in the DNA substrate and discover that they act as ``geometric catalysts'' for DNA integration. Additionally, we discover that the 1D and 3D clustering of loops affects the distribution of integration sites. Finally, we show that loops may compete with nucleosomes at attracting DNA integrations. These results may be tested \emph{in vitro} and they may help to understand patterns of DNA insertions with implications in genome evolution and gene therapy.
}
\end{abstract}

\maketitle

\section*{Introduction}
Genomes are constantly reshaped and manipulated during transcription, replication and cell division. They also need to be reshuffled during eukaryotic meiosis and undergo horizontal exchange of genetic material in prokaryotes. DNA transposition is one of the main factors driving genetic diversity and expansion, especially in plants. Transposons make up about 85\% of the maize genome~\cite{Kazazian2004,McClintock1950} and up about 50\% of the human genome~\cite{Lander2001} and can be associated with the onset of diseases~\cite{Payer2019}.
At the same time, the infection and replication of many viruses, including HIV-1, require the integration of the viral DNA within the host genome. Given the large abundance of non-transcribed DNA within the human genome (up to 90\%~\cite{cook2001principles}), a random insertion process would most of the time lead to unsuccessful infections. Instead, HIV-1 and other lentiviruses, appear to integrate their DNA non-randomly and in vicinity of transcriptionally active genes~\cite{Lucic2019a, Marini2015, Sultana2019}.
Additionally, both viral and transposable element insertions often appear clustered~\cite{Lucic2019a}, e.g. the LINE-1 and Alu repeats~\cite{Sultana2019}. These clusters of repeated elements often contribute to the organisation and compartmentalization of the genome, thereby impacting the global genome organisation~\cite{Schmidt2012,Sun2018,Grob2019,Jupe2019,Dong2018,Raviram2018,Cournac2016}. Viceversa, activation of transcription of transposable elements can cause the unfolding of chromatin and the downstream spatial rearrangement of the genome~\cite{Sun2020}.

Understanding the process of DNA insertion and the biophysical mechanisms driving non-random insertion patterns and clustered integrations is an important outstanding question~\cite{Vanderlinden2019,Prizak2022}. While it has been argued that DNA topology and chromatin structure have an impact on the integration site selection~\cite{Pasi2016,Michieletto2019,Bousios2020}, a systematic quantification of how integration is favoured or disfavoured in certain DNA and chromatin topologies remains an open challenge~\cite{Pruss1994, Benleulmi2015,Naughtin2015,Matysiak2017}.

Recent works have highlighted the importance of spatial and physical constraints in the integration patterns of HIV-1~\cite{Marini2015,Michieletto2019}. For instance, genes that are located closer to the nuclear envelope are more often integrated by HIV~\cite{Marini2015}. At the same time, insertion of transposable elements may be limited by the accessibility of transposase into chromatin~\cite{Prizak2022}, as employed in the ``assay for transposase-accessible chromatin'' (ATAC)~\cite{Buenrostro2013,Maeshima2015,Shin2018a}. Additionally, previous simulations have suggested that different physical features may be important at different scales during the process of integration~\cite{Michieletto2019}. For instance, HIV-1 viral DNA associated to the integrase enzyme first enters the nuclear envelope and diffuses through the nucleus within the large-scale chromatin mesh~\cite{Bosse2015,Burdick2020}; then, it likely binds and diffuses a chromatin fibre~\cite{Pruss1994a} and ultimately attempts to integrate within DNA~\cite{Jones2016a}, which requires elastic deformation of the DNA double-helix~\cite{Kvaratskhelia2014}. DNA integration is a multi-scale problem, involving a 3D search in a complex environment~\cite{Bosse2015,Marini2015} and a 1D search within a complex free energy landscape~\cite{Vanderlinden2019}. For this reason, in order to fully understand DNA integration, one requires to dissect the important contributions at each length-scale~\cite{Michieletto2019}.

In this work, we focus on the contribution of DNA looping, its 3D organisation, and competition with nucleosomes. Specifically, we first quantify how the length of DNA loops affects integration site selection and we discover that the longer the loops the less likely they are to attract integrations. Intriguingly, the rate of integration per unit length of loop is non-monotonic and displays a minimum: short loops are energetically favourable to integrate but difficult to find, while larger loops are less energetically favourable but more likely to be found and attract more integration attempts. We also find that the way loops are organised matters: clustered loops are systematically less integrated than sparse ones. Finally, we show that within a landscape where both loops and nucleosomes are present, loops are significantly more integrated if their length is shorter than the one of nucleosomal DNA.

\section*{Methods}
We model the insertion of a DNA element, e.g. transposons or viral DNA, into a DNA substrate, e.g. human or bacterial genome, using a coarse-grained bead-spring polymer model. Since integration is a relatively rare event that we want to study in isolation, using a coarse-grained model allows us to sample a far larger time and statistics than all-atom or oxDNA models. In our model, DNA is represented as a semi-flexible bead-spring chain polymer composed of $N$ beads of diameter $\sigma$, which is set to be $2.5$ nm (or $7.35$ bp). The dynamics of each bead are determined by a Langevin equation:
\begin{equation}
    m_i\frac{d^2\bm{r}_i}{dt^2} = -\bm{\nabla}U_i - \gamma_i\frac{d\bm{r}_i}{dt} + \delta\bm{F}_i 
\label{Langevineq}
\end{equation}
where $\bm{r}_i$ is the position of the i-th bead, $m_i$ and $\gamma_i$ are respectively the mass and the friction coefficient of the $i$-th bead due to an implicit solvent. Energies are expressed in units of $k_B T$, where $k_B$ is the Boltzmann constant and $T$ is the system's temperature. Distances are expressed in units of $\sigma$. Time is expressed in units of the typical time for a bead to diffuse a distance of its size $\tau_{Br}=\sigma^2/D = 3\pi\eta\sigma^3/k_BT$, where $D$ is the diffusion constant for a bead, and $\eta$ the viscosity of the implicit solvent. For simplicity, all the particles have the same mass and friction coefficient. The last term in Eq.~\eqref{Langevineq}, $\delta \bm{F}$, is a stochastic force with zero mean $\langle\delta\bm{F}(t)\rangle = \bm{0}$ and amplitude $\langle\delta F_{i,\alpha}(t)\delta F_{j,\beta}(t')\rangle = 2k_BT\,\gamma\,\delta_{ij}\,\delta_{\alpha\beta}\,\delta(t-t')$, where $i$ and $j$ run over particles and $\alpha$ and $\beta$ run over the Cartesian components. This choice ensures detailed balance~\cite{Frenkel2001}. The term containing the gradient $\bm{\nabla} U_i$ is the force acting on particle $i$ due to all the other particles in the system as explained below. Consecutive polymer beads are connected together through an harmonic potential:
\begin{equation}
    U_{harm} = k_{harm}(r_{i,i+1}-r_0)^2
    \label{harmeq}
\end{equation}
Where $k_{harm}$ (set to $20k_BT/\sigma^2$) determines the strength of the spring and $r_0$ (set to $1.1\sigma$) is the equilibrium bond distance. We implement the bending rigidity of the DNA through a Kratky-Porod potential between triplets of consecutive beads
\begin{equation}
    U_{bend} = \frac{k_BTl_p}{\sigma}\left[1-\frac{\bold{t}_i\cdot\bold{t}_{i+1}}{|\bold{t}_i||\bold{t}_{i+1}|}\right]
    \label{bendeq}
\end{equation}
where $\bold{t}_i$ is the tangent vector connecting beads $i-1$ to $i$, and $l_p$ is the persistence length of the polymer which is set to be $l_p=20\sigma=50$ nm. Finally, steric hindrance interactions between non-adjacent polymer beads are simulated through a LJ potential, shifted to be zero at $r_{cut}=2^{\frac{1}{6}}\sigma$, as
\begin{equation} \label{eqLJcut}
    U_{LJ,cut} = U_{LJ}(r_{i,j}) - U_{LJ}(r_{cut})
\end{equation}
if $r < r_{cut}$ and 0 otherwise, where
\begin{equation}
    U_{LJ}  = 4\epsilon \left[ \left( \dfrac{\sigma}{r_{i,j}} \right)^{12} - \left( \dfrac{\sigma}{r_{i,j}} \right)^{6} \right] \, .
\end{equation}
The total potential energy acting on bead $i$ is the sum of these three contributions. Equation~\eqref{Langevineq} is integrated using the Large-scale Atomic/Molecular Massively Parallel Simulator (LAMMPS) with a velocity–Verlet scheme~\cite{Plimpton1995}. We set the integration time step to be $\Delta t = 0.01\tau_{Br}$. Our model does not account for the torsional rigidity of the DNA, and we plan to include it in the future.

\begin{figure}[t!]
\centering
\includegraphics[width=.45\textwidth]{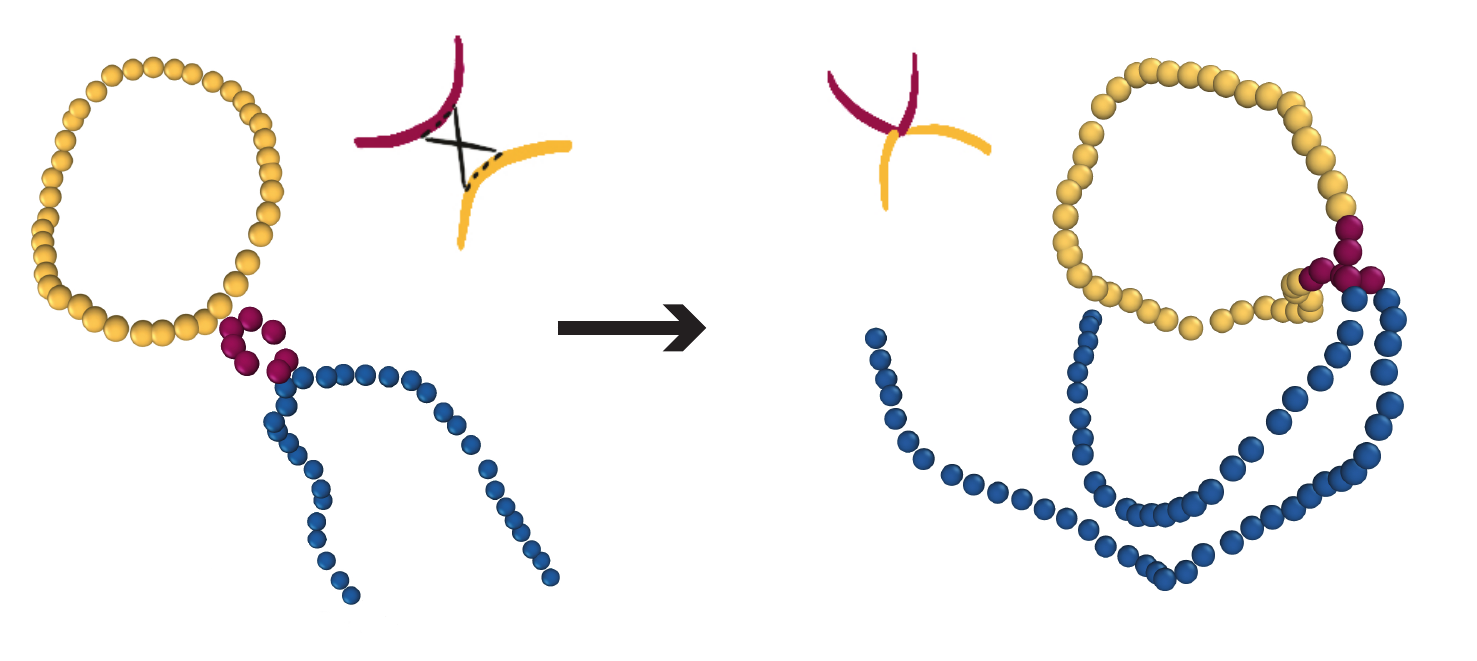}
\caption{Snapshot from molecular dynamics simulations displaying an integration event within a DNA loop. Blue beads represent non-looped target DNA, purple beads looped target DNA and yellow beads represent viral DNA.}
\label{modelfig}
\end{figure}

Using the model described above we simulate two chains, one circular representing viral DNA or a transposable element, and the other linear representing the target, or substrate DNA. Both polymers diffuse in a box and we attempt the integration of the circular polymer into the linear chain using a modified version of the ``double-bridging'' algorithm implemented in LAMMPS as \verb|fix bond/swap|~\cite{Sides2004}. This code allows us to perform ``reconnection'' moves and swap the bonds in the system that are connecting beads that are closer than $R_c = 2 \sigma$ in 3D (see Fig.~\ref{modelfig}). The reconnections are attempted every 1 LAMMPS step. We implemented some modifications to this fix that allow us to perform recombination moves between viral and host polymers (inter-chain reconnections) while avoiding intra-chain (or self) reconnections~\cite{Michieletto2019,forte2021investigating,Bonato2022}. The reconnection moves are also weighted by a Metropolis test that assigns a probability of successful swap depending on the energy difference $\Delta U$ between the old and the new configuration before and after the swap. This probability is 1 if $\Delta U < 0$ and $p = e^{-\Delta U/k_BT}$ if $\Delta U \geq 0$. We also modified this part of the code to perform polymer reconnections that bypass the Metropolis test thus allowing non-equilibrium integration. These modified codes can be found at \url{https://git.ecdf.ed.ac.uk/taplab} .

\begin{figure*}[t!]
\centering
\includegraphics[width=.95\textwidth]{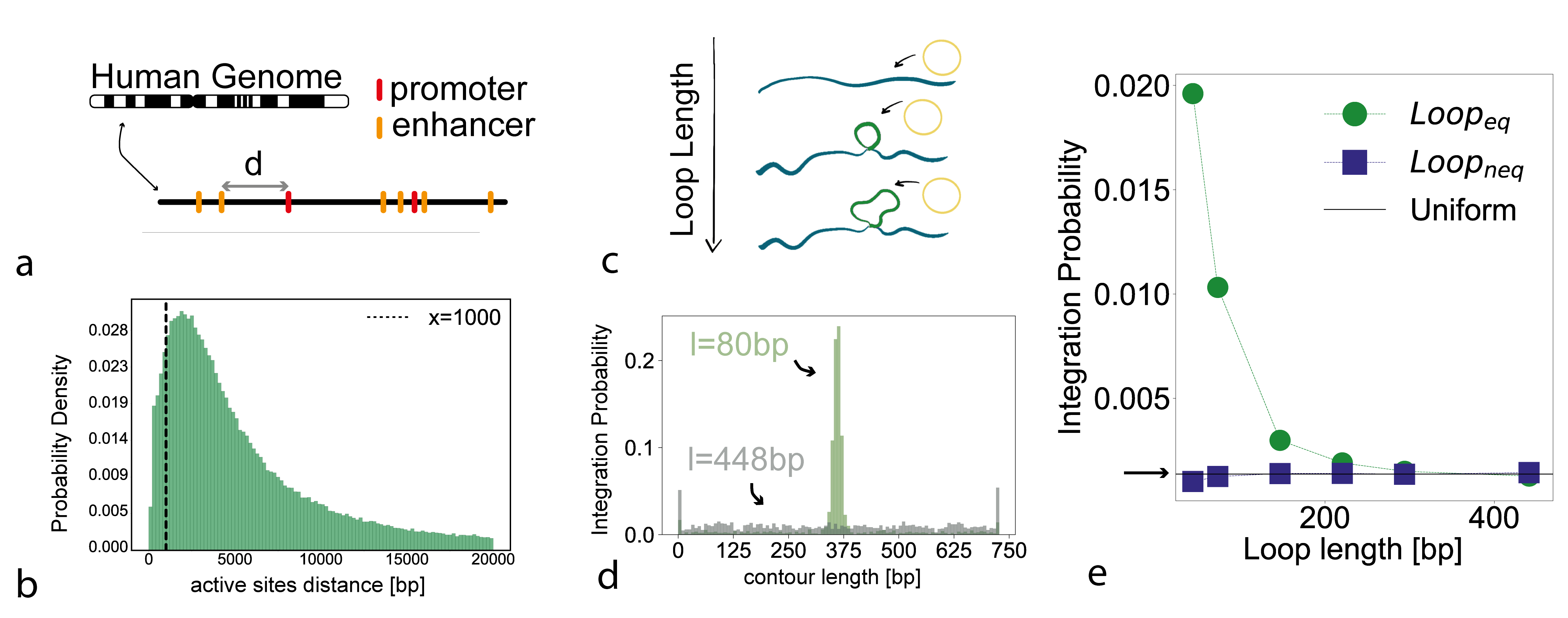}
\caption{\textbf{a} Sketch of our search procedure for nearest 1D regulatory elements in the genome \textbf{b} Histogram of 1D distances between nearest-neighbor pairs of regulatory elements, i.e. enhancer and promoters, within the human genome: $3\%$ of these are below $500$ bp and $8\%$ below $1$ kbp~\cite{Kent2002}. \textbf{c} Sketch of the simulation setup, where we investigate the integration probability in loops of different sizes. \textbf{d} Histogram of integration sites probability along the polymer for loops of different sizes (80 bp in green and 448 bp in grey).  \textbf{e} Probability of integration inside the looped region as a function of loop length, comparing equilibrium (green) and non-equilibrium (purple) algorithms. Random integration would yield the black line indicated by the arrow.
}
\label{loops1d}
\end{figure*}

The simulations are performed as follows. We prepare an initial polymer configuration as a random walk and impose the loop by setting a harmonic bond (with strength increasing from 1 to 20 $k_BT$ during the equilibration) in between two beads at 1D distance $l$. For each of the $>$ 1000 replicas, we let the system equilibrate for at least $3 \, 10^5$ LAMMPS steps. At the end of the equilibration, we allow the integration to happen. We stop the simulation after $10^7$ timesteps, irrespectively of whether the integration has happened or not. We output the bond list every $10^6$ steps and from that we reconstruct the topology of the polymers and can detect when and where the integration has happened. Using this information we then build a histogram of where the integration sites are along the polymer and can thus quantify how many are within and outside the looped regions.

\section*{Results}

\subsection*{Integration in DNA loops}

DNA loops are thought to be abundant and biologically important in both eukaryotic and prokaryotic genomes~\cite{Finzi1995,Priest2014,Ding2014,Rao2014,Yan2018}. Beyond classic enhancer-promoter interactions~\cite{Alberts2014}, several new mechanisms of loop formation have been identified recently, for instance those mediated by CTCF~\cite{Rao2014} ($\sim$ 0.1-1 Mbp), cohesin~\cite{Ryu2021,Davidson2019} and condensin~\cite{Hirano2012,Ganji2018} ($\sim$ 1-10 kbp). In bacteria, structural-maintenance-of-chromosome (SMC) proteins~\cite{Brandao2019a,Brandao2021} and nucleoid associated proteins~\cite{Yoshua2021} also play a major role in forming DNA loops \emph{in vivo}~\cite{Le2013}(30-420 kbp). On top of this, integrase can itself deform the substrate in such a way to create a small ($\sim$ 5 - 10 bp) DNA loop within the nucleosome~\cite{Wilson2019}. At the same time, investigation of DNA loops in vitro is also possible~\cite{Finzi1995,Brouns2018}.

In analogy with previous work, here we argue that short DNA loops may attract more integrations due to the fact that they have a lower energy barrier for elastic deformations~\cite{Michieletto2019}. On the other hand, short DNA loops are energetically costly~\cite{Ryu2021,Sankararaman2005,Brahmachari2018}. In order to quantify how frequently short DNA loops may appear in the human genome, we compute the 1D distance between nearest-neighbouring regulatory elements, such as enhancers and promoters, as listed in the GenHancher database~\cite{Kent2002}. Fig.~\ref{loops1d}a shows that about 8\% of these regulatory elements are less than 1 kbp apart and 3\% of them shorter than 500 bp. While only a fraction of them will be looped at any one time, these figures show that there is a considerable amount of short ($< 1$ kbp) potential loops in the human genome. On top of this, we note that the integrase itself can compete with nucleosomes to create an extremely short loop of DNA in the range of 5-10 bp~\cite{Wilson2019}. Likewise, we expect similar, if not greater proportion of short loops between regulatory elements in bacteria, such as the lac repressor~\cite{Kramer1987} or deformations of the DNA substrate for instance created by nucleoid-associated proteins such as IHF~\cite{Yoshua2021,Fosado2023}.

Motivated by this, we aim to characterise how the 1D loop length affects the insertion of the circular DNA into the target substrate DNA. To do this, we perform simulations in which a loop of length $\ell$ is stabilised in a polymer of length $N=100$ beads $=735$ bp introducing a harmonic spring in between two beads. We then perform 1500 independent simulations to stochastically sample the probability distribution, $p(x)$, of integrating at position $x$ along the polymer. In Fig.~\ref{loops1d}d we show the profile of integrations, i.e. $p(x)$, over the simulated polymer. One can appreciate that the case in which there is a short loop (80 bp) in the polymer yields a distribution $p(x)$ peaked within the loop, whereas the case with a longer loop (448 bp) yields a constant distribution compatible with the random $p_{rand} = 1/N$ distribution.

To better quantify the enhancement of integrations within the looped region compared with the ones outside it, we compute the probability of integration per bead inside the loop as $p_{in} = I_{in}/(\ell I_{tot})$, where $I_{in}$ is the number of integrations that occur inside the loop and $I_{tot}$ the total number of integrations across our simulations. If the insertions were random we would expect $I_{in}/I_{tot}=\ell/N$, recovering a random integration probability \emph{per bead} $p_{rand} = 1/N$.

\begin{figure}[t!]
\centering
\includegraphics[width=0.48\textwidth]{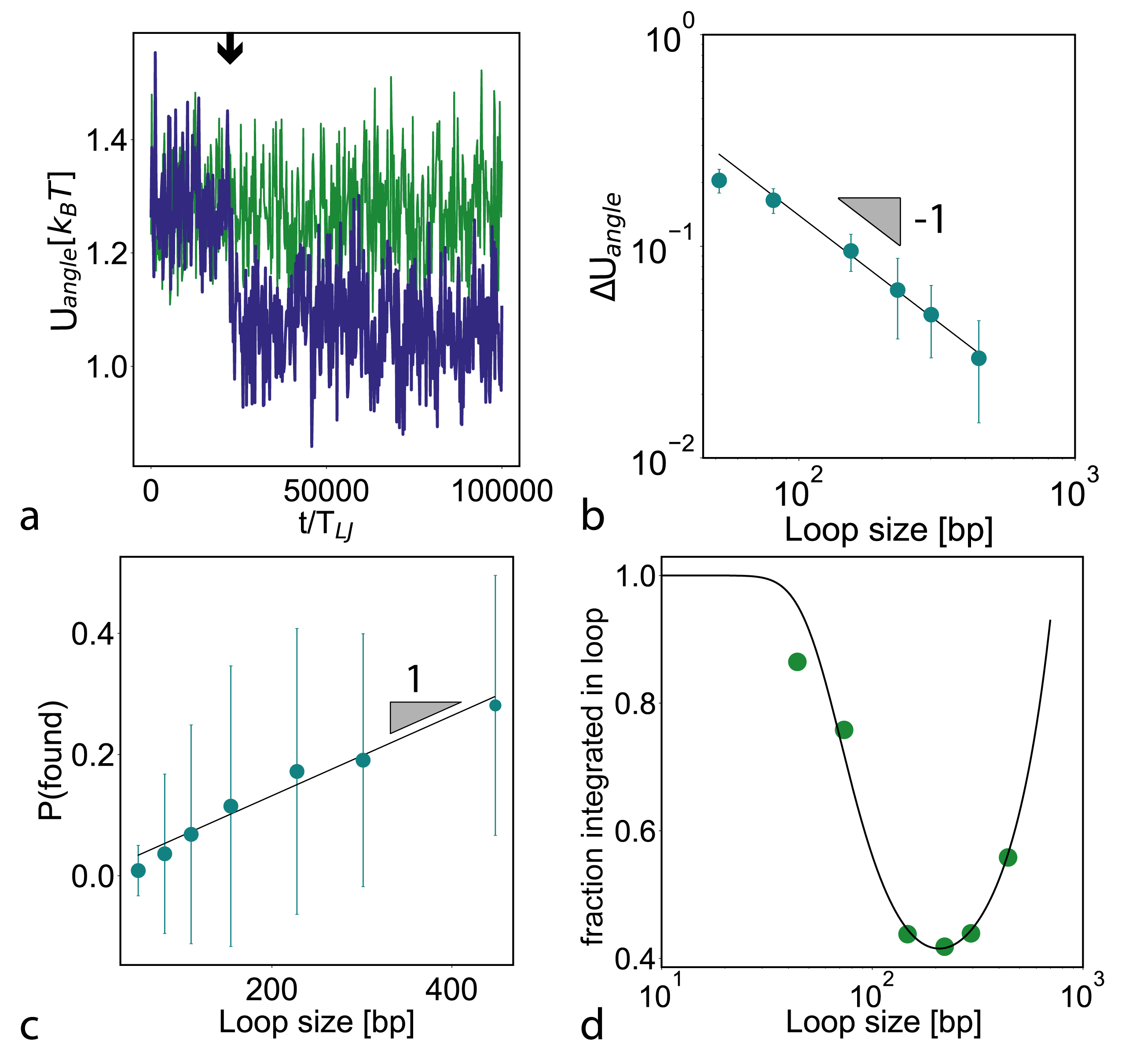}
\vspace{-0.5 cm}
\caption{\textbf{a} Angle energy measured during the course of two simulations, one displaying an integration event within a loop of length 51 bp (blue trajectory) and one displaying an integration in the non-looped region of the polymer (green).  \textbf{b} Scaling of the change in (free) energy before/after an integration event within a loop as a function of the loop size, displaying a $\Delta U \sim 1/\ell$ dependence. \textbf{c} Scaling of the fraction of time the viral DNA spends in the 3D vicinity of the looped section as a function of loop length and displaying a linear scaling. \textbf{d} Fraction of integrations within the looped region. The black line is a fit of Eq.~\eqref{eq:f_in} with $\Delta U/k_B T$ and $b$ as free parameters.}
\vspace{-0.5 cm}
\label{loops1dtheory}
\end{figure}

\begin{figure*}[t!]
\centering
\includegraphics[width=.9\textwidth]{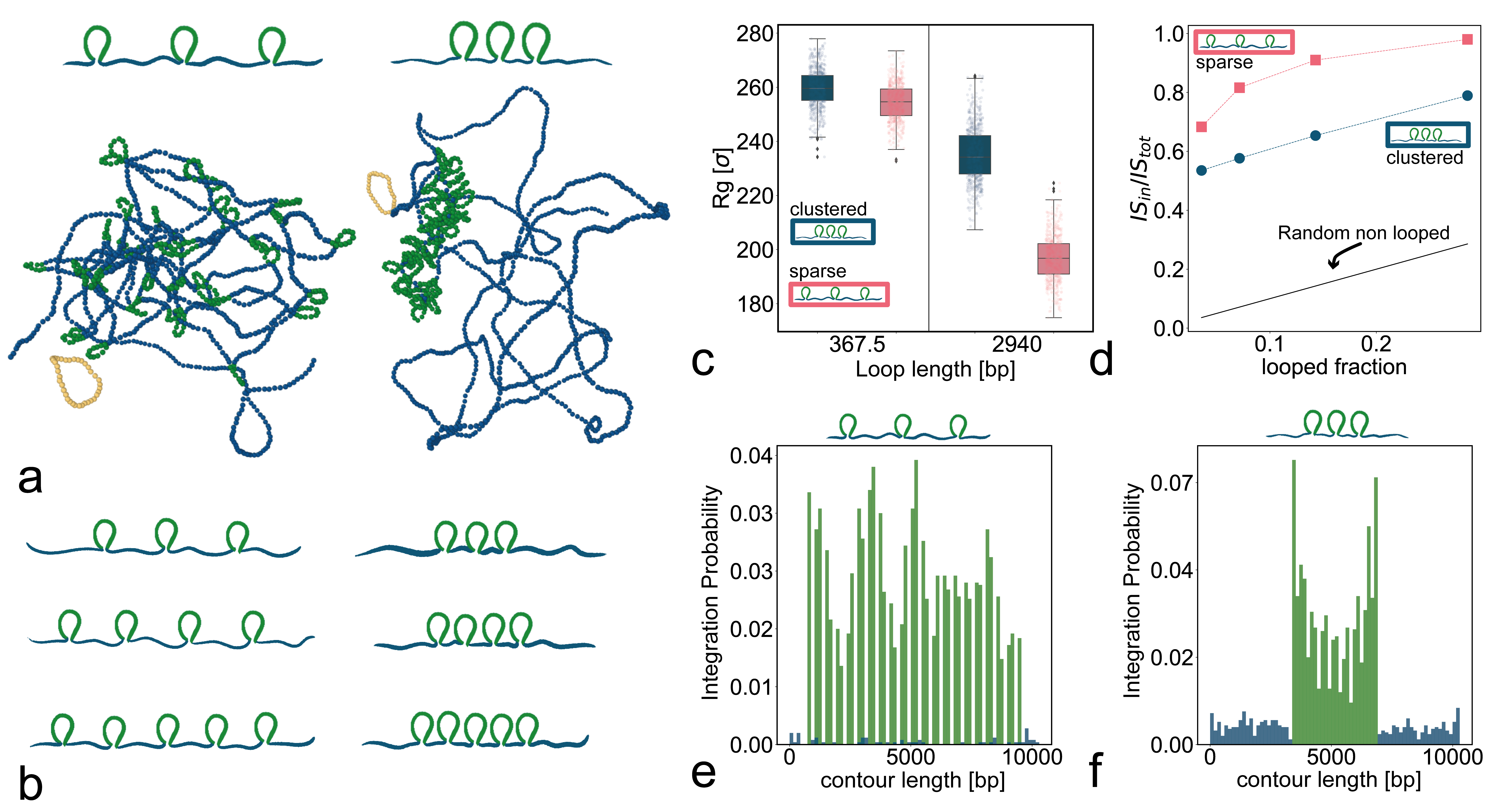}
\caption{\textbf{a} Sketch and snapshots of our simulated setup. One can appreciate that sparse loops cover a larger volume fraction than clustered ones. \textbf{b} We keep the length of the loops constant to $\ell=80$ bp and vary the number of loops on the polymer, increasing the fraction of contour length covered by loops. \textbf{c} Radius of gyration of the region hosting loops. \textbf{d} Fraction of integrations inside loops normalised by the total amount of integrations. Polymers with clustered loops display a systematically smaller fraction of integrations occurring in loops. \textbf{e-f} Histogram of integration probability along the polymer in the case of (e) sparse and (f) clustered, loops. }
\label{loop3d}
\end{figure*}

By analysing our simulations, we find that $p_{in}$ is significantly larger than the random probability when the polymer has a short loop ($<$ 200 bp), and that $p_{in}$ decays to $1/N$ for large loop lengths. We understand this as follows. Short loops store a significant amount of energy in the form of looping. This implies that when the viral loop integrates into the pre-looped substrate, it lowers the (free) energy of the system because the loop is then longer and the curvature smaller. This is also in line with the experimental finding of favoured integration in DNA molecules carrying torsional stress~\cite{Vanderlinden2019} and can be explained as follows: the elastic energy paid for looping of an elastic rod of length $\ell$ and persistence length $l_p$ is
\begin{equation}
 \dfrac{F}{k_BT} = \dfrac{\epsilon l_p}{\ell}
 \label{eq:freeen}
\end{equation}
with $\epsilon$ a numerical factor depending on the shape of the loop~\cite{Dittmore2017}. Assuming the integration process to be an Arrhenius, energy activated, process we expect the integration in looped substrates to occur at a rate $\kappa = \exp{\Delta U/k_BT}$ where $\Delta U = U(\textrm{after}) - U(\textrm{before}) < 0$. In figure (Fig.~\ref{loops1dtheory}a, blue) we indeed show that an integration event within a loop (indicated with an arrow) lowers the energy of the system. On the contrary, for integrations in non-looped substrates $\Delta E_{out}=0$, as there is no free energy gain (Fig.~\ref{loops1dtheory}a, green). We can compute the energy difference before/after integration within a loop as $\Delta U = U(\textrm{after})- U(\textrm{before}) \approx a/(\ell + l_{HIV}) - a/\ell \approx -a/\ell + const$ for  $\ell \ll l_{HIV}$ and where $a$ is a numerical constant. We test this approximation by measuring the energy change during an integration event and we plot the difference $\Delta U$ in Fig.~\ref{loops1dtheory}b which indeed displays a $1/\ell$ dependence, as expected.
In turn, the probability of integration within a loop can be written as
\begin{equation}
P(\textrm{found} | \textrm{in} ) P(\textrm{in}) = P(\textrm{in} | \textrm{found}) P(\textrm{found})
\end{equation}
and where we can take $P(\textrm{found} | \textrm{in} ) = 1$. In this equation $P(\textrm{found})$ is the probability of a viral DNA to find a loop, and $P(\textrm{in})$ is the probability of integration in the loop. In turn, we can write that the probability that there is an integration within the loop, conditional to the fact that the viral DNA has found the loop is $P(\textrm{in} | \textrm{found})= \exp{(-\Delta U/k_BT)} \sim \exp{(- a/\ell)}$. At the same time, the probability that the viral DNA finds the looped segment within the simulation box is $P(\textrm{found}) = 4/3 \pi \sigma^3 \ell/V \sim \ell$, where $V$ is the volume of the box. In simulations, we can test this scaling by tracking the fraction of time that the viral loop spends in 3D proximity of the DNA loop as a function of different loop sizes; as shown in Fig.~\ref{loops1dtheory}c, this quantity is to a good extent linearly increasing with the size of the loop. The same argument holds for integrations outside of the loop, where $P(\textrm{out} | \textrm{found}) = \exp{(-\Delta U/k_BT)} \sim 1$ (as there is no energy difference) and $P(\textrm{found}) \sim (N - \ell)$. Finally, we can write that the fraction of viral loops integrated within the looped substrate is then expected to be
\begin{equation}
f_{in} = \dfrac{P(\textrm{in})}{P(\textrm{in}) + P(\textrm{out})} = \dfrac{\ell e^{- a/\ell + b}}{\ell e^{- a/\ell + b} + (N - l)}  \, .
\label{eq:f_in}
\end{equation}
where $a$ is expected to be proportional to $\Delta U/k_BT$ and $b$ a constant. By plotting the fraction of integrations inside the looped region $f_{in}= p_{in} \ell$ we see that the data follows extremely well the prediction of Eq.~\eqref{eq:f_in} (see Fig.~\ref{loops1dtheory}d).
Intriguingly, our findings imply that there is an optimum length to reduce the amount of integrations within the looped segment. As one can appreciate from Fig.~\ref{loops1dtheory}d, the minimum corresponds to a loop that is long enough to possess a small free energy gain upon integration, yet short enough to be difficult to find.

\subsection*{Integration in clustered and sparse loops}

We now ask what happens to the distribution of integrations when there are many loops along the substrate. In particular, we are interested to understand what are the cooperative effects that appear when the loops are sparse and uniformly distributed along the substrate or when they are clustered in a short segment of the polymer (Fig.~\ref{loop3d}a). To investigate this question, we therefore perform simulations with many loops (all of the same length $\ell=80$ bp) formed along a polymer of 10.3 kbp in either sparse or clustered 1D arrangement (Fig.~\ref{loop3d}a).

Interestingly, we observe that the 3D conformations of the polymers are rather distinct, with the latter (clustered) case being more swollen and with the looped region occupying a smaller volume fraction than in the former (sparse) case (see snapshots in Fig.~\ref{loop3d}a and radius of gyration in Fig.~\ref{loop3d}c).

To then understand if this distinct 3D organisation due to the 1D arrangement affects the overall integration probability, we perform simulations with varying numbers of loops while preserving constant loop size (Fig.~\ref{loop3d}d). In line with what we discovered in the previous section, we observe that the sparse loops attract more integrations overall. Again, we argue that this is due to the smaller 3D volume fraction occupied by the 1d clustered loop arrangement.

The consequence of such a different 3D organisation is reflected on the distribution of integration events. Indeed, as shown in Fig.~\ref{loop3d}b-c, we observe that in the sparse case the integrations are uniformly distributed among all the loops in the polymer, however in the case of clustered loops, the ones at the two ends of the region hosting the loops display more integrations than the inner ones (Fig.~\ref{loop3d}c). We argue that this is due to a ``screening'' effect, whereby the marginal loops are typically the ones more exposed while the more internal ones are more tucked in within the clustered looped region.

\subsection*{Integration in looped euchromatin}
We conclude this work by investigating the integration within a substrate that displays both loops and nucleosomes. Nucleosomes are assembled as in Ref.~\cite{Michieletto2019} where we have previously shown that DNA wrapped around histones attract more integrations due to their strongly bent states, in line with experiments~\cite{Vanderlinden2019, Wilson2019, Pruss1994, Pruss1994a}. Here we ask what happens if in a segment of chromatin there is a competition, or a synergy, between looped and nucleosomal DNA and whether the two would enhance even more the integration in looped euchromatin.

\begin{figure}[t!]
\centering
\includegraphics[width=0.5\textwidth]{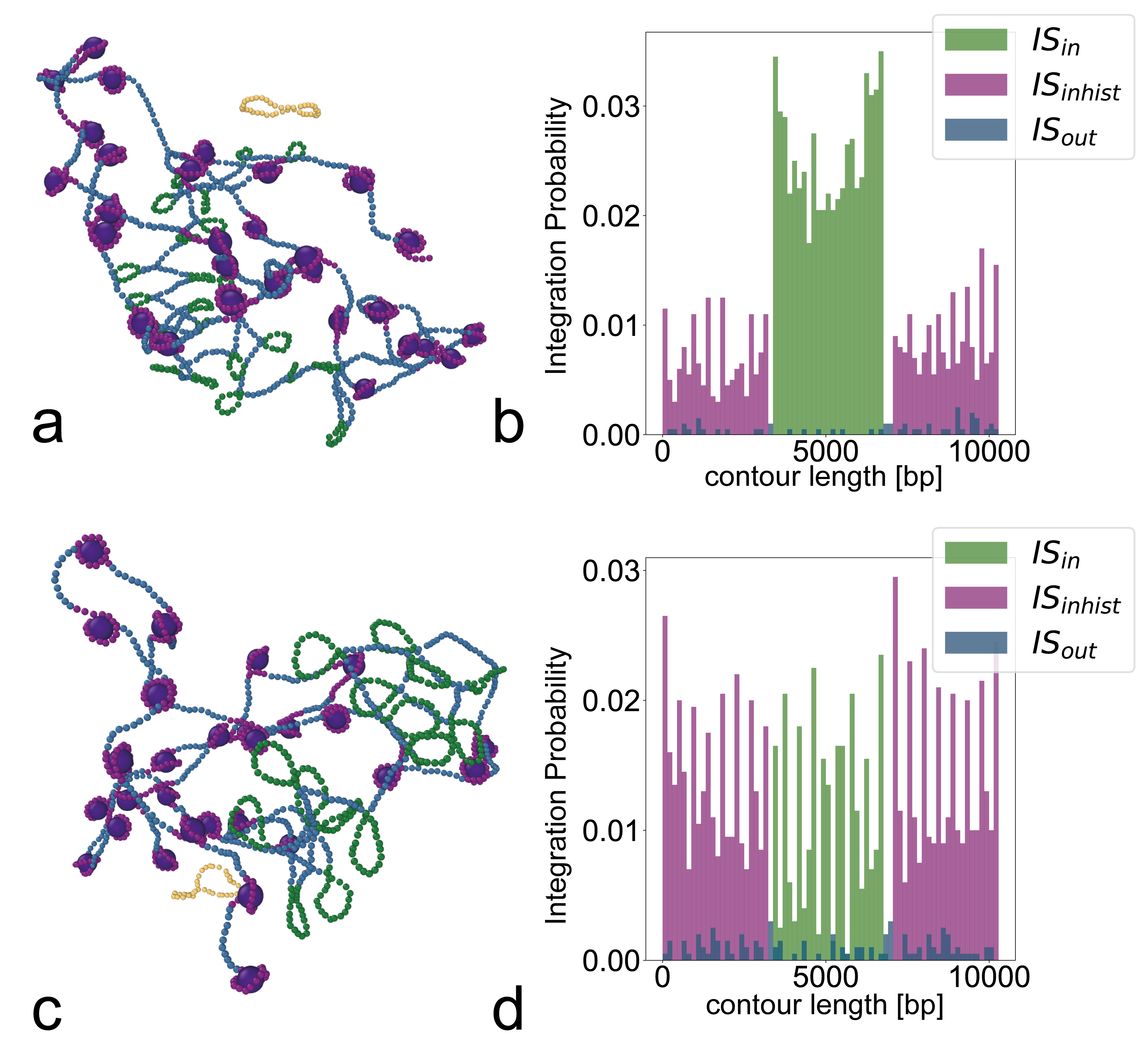}
\caption{\textbf{a} Snapshot of chromatin simulation with short $l=80$ bp loops. \textbf{b} Probability of integration along the polymer with short loops. \textbf{c} Snapshot of chromatin simulation with loops comparable with nucleosome length $l=154$ bp. \textbf{d} Probability of integration along the polymer with long loops. }
\label{euheterofig}
\end{figure}

To test this, we perform simulations of a polymer of length 10.3 kbp wrapped around 30 histones (where each block of nucleosomal DNA is 20 beads or $\sim$ 147 bp) and displaying a central looped region with varying loop lengths. Specifically, we compare two systems: one with loops of length $\ell = 80$ bp (11 beads), and one with loops of length $\ell = 154$ bp (21 beads)  (see snapshots in Fig.~\ref{euheterofig}a,c). We observe that shorter loops enhance integration within the looped region rather than into the nucleosomal DNA. Instead, having loops of approximately the same size as the nucleosomal DNA blocks leads to a more uniformly distributed integration profile, preserving only the preference of ``bent'' DNA regions over the rest of the polymer (Fig.~\ref{euheterofig}b,d). We note that the spikes in the integration profile correspond to the location of a nucleosome or a loop. In line with what we discovered in the previous section, we again observe in the profile of integration probability, that the interior loops and nucleosomes are screened by the more exterior ones (i.e. $P(s)$ is systematically smaller in the middle with respect to the edge of the polymer).

Finally, we highlight that recent cryo-EM structural data revealed that the intasome-nucleosome structure creates a short, highly bent DNA loop by shifting the nucleosome out of registry~\cite{Wilson2019}. Such a short (5-10 bp) loop would, in our model, favour even more the integration within this extremely bent loop-on-nucleosome feature.

\section{Conclusions}
How geometric features of DNA and chromatin may affect the integration site selection of retroviral DNA or transposable elements is far from understood.  While it is widely accepted that the integration site selection is not random, the underpinning (bio)physical processes driving the selection are still largely unexplored.

In this work, we used computer simulations to explore the role of loops in DNA and chromatin in determining integration site selection. We use a model of reconnecting polymers in equilibrium, hence iso-energetic and satisfying detailed balance, that has already been successful at capturing the distribution of HIV integration site selection in the human genome~\cite{Michieletto2019}.
First, we discover that integrations are favoured in looped regions due to the release of bending energy. This is in line with experimental observations showing that integration is favoured in nucleosomes~\cite{Pruss1994} and supercoiled substrates~\cite{Vanderlinden2019}. In spite of the fact that the microscopic action of strand exchange is iso-energetic, the release of bending energy acts as a geometric catalyst for the integration site selection.

Intriguingly, we observe a competition between the free energy gained and the probability of finding such loops. More specifically, integration in small loops releases a larger amount of free energy compared with larger loops, but they are more difficult to find and vice-versa. In our simulations, this competition manifests itself with a non-monotonic behaviour of the integration rate displaying a minimum around 200 bp.

We have also explored the effect of having many loops on the same substrate. Interestingly, the 3D conformation of the polymer strongly depends on whether the loops are sparse, i.e. uniformly distributed along the chain, or are clustered in 1D. In the latter case, the looped region occupies a smaller fraction of the simulation box and is, as a consequence, overall less integrated than the sparse loops.  This many-loop system also displays a screening effect whereby 1D clustered loops display a non-uniform integration probability, that is largest for the loops that are most external to the cluster.

Finally, we perform simulations of a chromatin substrate and discover that loops attract more integrations, as long as they are shorter than the nucleosomal DNA segments. Our simulations suggest that loops-on-nucleosomes, such as the ones generated by the intasome-nucleosome complex~\cite{Wilson2019}, are perhaps the most effective at attracting integrations.

We argue that our results will illuminate the role of loops as geometric catalysts in determining the integration site selection \emph{in vivo}. We argue that our results could be tested by designing artificial loops either \emph{in vivo} using, for instance, bivalent dCas9~\cite{Hao2017} or \emph{in vitro} DNA origami~\cite{Rothemund2006} or generic bridge proteins~\cite{Ryu2021}.

\acknowledgements{
DM is a Royal Society University Research Fellow and is supported by the ERC (Starting Grant, Topologically Active Polymers, Ref.~947918).
Source codes are available at \url{https://git.ecdf.ed.ac.uk/taplab}.
}

\bibliographystyle{apsrev4-1}
\bibliography{library}

\begin{thebibliography}{62}%
\makeatletter
\providecommand \@ifxundefined [1]{%
 \@ifx{#1\undefined}
}%
\providecommand \@ifnum [1]{%
 \ifnum #1\expandafter \@firstoftwo
 \else \expandafter \@secondoftwo
 \fi
}%
\providecommand \@ifx [1]{%
 \ifx #1\expandafter \@firstoftwo
 \else \expandafter \@secondoftwo
 \fi
}%
\providecommand \natexlab [1]{#1}%
\providecommand \enquote  [1]{``#1''}%
\providecommand \bibnamefont  [1]{#1}%
\providecommand \bibfnamefont [1]{#1}%
\providecommand \citenamefont [1]{#1}%
\providecommand \href@noop [0]{\@secondoftwo}%
\providecommand \href [0]{\begingroup \@sanitize@url \@href}%
\providecommand \@href[1]{\@@startlink{#1}\@@href}%
\providecommand \@@href[1]{\endgroup#1\@@endlink}%
\providecommand \@sanitize@url [0]{\catcode `\\12\catcode `\$12\catcode
  `\&12\catcode `\#12\catcode `\^12\catcode `\_12\catcode `\%12\relax}%
\providecommand \@@startlink[1]{}%
\providecommand \@@endlink[0]{}%
\providecommand \url  [0]{\begingroup\@sanitize@url \@url }%
\providecommand \@url [1]{\endgroup\@href {#1}{\urlprefix }}%
\providecommand \urlprefix  [0]{URL }%
\providecommand \Eprint [0]{\href }%
\providecommand \doibase [0]{http://dx.doi.org/}%
\providecommand \selectlanguage [0]{\@gobble}%
\providecommand \bibinfo  [0]{\@secondoftwo}%
\providecommand \bibfield  [0]{\@secondoftwo}%
\providecommand \translation [1]{[#1]}%
\providecommand \BibitemOpen [0]{}%
\providecommand \bibitemStop [0]{}%
\providecommand \bibitemNoStop [0]{.\EOS\space}%
\providecommand \EOS [0]{\spacefactor3000\relax}%
\providecommand \BibitemShut  [1]{\csname bibitem#1\endcsname}%
\let\auto@bib@innerbib\@empty
\bibitem [{\citenamefont {Kazazian}(2004)}]{Kazazian2004}%
  \BibitemOpen
  \bibfield  {author} {\bibinfo {author} {\bibfnamefont {H.~H.}\ \bibnamefont
  {Kazazian}},\ }\href@noop {} {\bibfield  {journal} {\bibinfo  {journal}
  {Science}\ }\textbf {\bibinfo {volume} {303}},\ \bibinfo {pages} {1626}
  (\bibinfo {year} {2004})}\BibitemShut {NoStop}%
\bibitem [{\citenamefont {McClintock}(1950)}]{McClintock1950}%
  \BibitemOpen
  \bibfield  {author} {\bibinfo {author} {\bibfnamefont {B.}~\bibnamefont
  {McClintock}},\ }\href@noop {} {\bibfield  {journal} {\bibinfo  {journal}
  {Proceedings of the National Academy of Sciences of the United States of
  America}\ }\textbf {\bibinfo {volume} {36}},\ \bibinfo {pages} {337}
  (\bibinfo {year} {1950})}\BibitemShut {NoStop}%
\bibitem [{\citenamefont {Lander}\ \emph {et~al.}(2001)\citenamefont {Lander}
  \emph {et~al.}}]{Lander2001}%
  \BibitemOpen
  \bibfield  {author} {\bibinfo {author} {\bibfnamefont {E.~S.}\ \bibnamefont
  {Lander}} \emph {et~al.},\ }\href@noop {} {\bibfield  {journal} {\bibinfo
  {journal} {Nature}\ }\textbf {\bibinfo {volume} {412}},\ \bibinfo {pages}
  {565} (\bibinfo {year} {2001})}\BibitemShut {NoStop}%
\bibitem [{\citenamefont {Payer}\ and\ \citenamefont
  {Burns}(2019)}]{Payer2019}%
  \BibitemOpen
  \bibfield  {author} {\bibinfo {author} {\bibfnamefont {L.~M.}\ \bibnamefont
  {Payer}}\ and\ \bibinfo {author} {\bibfnamefont {K.~H.}\ \bibnamefont
  {Burns}},\ }\href@noop {} {\bibfield  {journal} {\bibinfo  {journal} {Nature
  Reviews Genetics}\ }\textbf {\bibinfo {volume} {20}},\ \bibinfo {pages} {760}
  (\bibinfo {year} {2019})}\BibitemShut {NoStop}%
\bibitem [{\citenamefont {Cook}(2001)}]{cook2001principles}%
  \BibitemOpen
  \bibfield  {author} {\bibinfo {author} {\bibfnamefont {P.}~\bibnamefont
  {Cook}},\ }\href@noop {} {\emph {\bibinfo {title} {Principles of Nuclear
  Structure and Function}}},\ \bibinfo {series} {Principles of Nuclear
  Structure and Function}\ No.\ \bibinfo {number} {p. 954}\ (\bibinfo
  {publisher} {Wiley},\ \bibinfo {year} {2001})\BibitemShut {NoStop}%
\bibitem [{\citenamefont {Lucic}\ \emph {et~al.}(2019)\citenamefont {Lucic},
  \citenamefont {Chen}, \citenamefont {Kuzman}, \citenamefont {Zorita},
  \citenamefont {Wegner}, \citenamefont {Minneker}, \citenamefont {Wang},
  \citenamefont {Fronza}, \citenamefont {Laufs}, \citenamefont {Schmidt},
  \citenamefont {Stadhouders}, \citenamefont {Roukos}, \citenamefont
  {Vlahovicek}, \citenamefont {Filion},\ and\ \citenamefont
  {Lusic}}]{Lucic2019a}%
  \BibitemOpen
  \bibfield  {author} {\bibinfo {author} {\bibfnamefont {B.}~\bibnamefont
  {Lucic}}, \bibinfo {author} {\bibfnamefont {H.~C.}\ \bibnamefont {Chen}},
  \bibinfo {author} {\bibfnamefont {M.}~\bibnamefont {Kuzman}}, \bibinfo
  {author} {\bibfnamefont {E.}~\bibnamefont {Zorita}}, \bibinfo {author}
  {\bibfnamefont {J.}~\bibnamefont {Wegner}}, \bibinfo {author} {\bibfnamefont
  {V.}~\bibnamefont {Minneker}}, \bibinfo {author} {\bibfnamefont
  {W.}~\bibnamefont {Wang}}, \bibinfo {author} {\bibfnamefont {R.}~\bibnamefont
  {Fronza}}, \bibinfo {author} {\bibfnamefont {S.}~\bibnamefont {Laufs}},
  \bibinfo {author} {\bibfnamefont {M.}~\bibnamefont {Schmidt}}, \bibinfo
  {author} {\bibfnamefont {R.}~\bibnamefont {Stadhouders}}, \bibinfo {author}
  {\bibfnamefont {V.}~\bibnamefont {Roukos}}, \bibinfo {author} {\bibfnamefont
  {K.}~\bibnamefont {Vlahovicek}}, \bibinfo {author} {\bibfnamefont {G.~J.}\
  \bibnamefont {Filion}}, \ and\ \bibinfo {author} {\bibfnamefont
  {M.}~\bibnamefont {Lusic}},\ }\href {\doibase 10.1038/s41467-019-12046-3}
  {\bibfield  {journal} {\bibinfo  {journal} {Nature Communications}\ }\textbf
  {\bibinfo {volume} {10}} (\bibinfo {year} {2019}),\
  10.1038/s41467-019-12046-3}\BibitemShut {NoStop}%
\bibitem [{\citenamefont {Marini}\ \emph {et~al.}(2015)\citenamefont {Marini},
  \citenamefont {Kertesz-Farkas}, \citenamefont {Ali}, \citenamefont {Lucic},
  \citenamefont {Lisek}, \citenamefont {Manganaro}, \citenamefont {Pongor},
  \citenamefont {Luzzati}, \citenamefont {Recchia}, \citenamefont {Mavilio},
  \citenamefont {Giacca},\ and\ \citenamefont {Lusic}}]{Marini2015}%
  \BibitemOpen
  \bibfield  {author} {\bibinfo {author} {\bibfnamefont {B.}~\bibnamefont
  {Marini}}, \bibinfo {author} {\bibfnamefont {A.}~\bibnamefont
  {Kertesz-Farkas}}, \bibinfo {author} {\bibfnamefont {H.}~\bibnamefont {Ali}},
  \bibinfo {author} {\bibfnamefont {B.}~\bibnamefont {Lucic}}, \bibinfo
  {author} {\bibfnamefont {K.}~\bibnamefont {Lisek}}, \bibinfo {author}
  {\bibfnamefont {L.}~\bibnamefont {Manganaro}}, \bibinfo {author}
  {\bibfnamefont {S.}~\bibnamefont {Pongor}}, \bibinfo {author} {\bibfnamefont
  {R.}~\bibnamefont {Luzzati}}, \bibinfo {author} {\bibfnamefont
  {A.}~\bibnamefont {Recchia}}, \bibinfo {author} {\bibfnamefont
  {F.}~\bibnamefont {Mavilio}}, \bibinfo {author} {\bibfnamefont
  {M.}~\bibnamefont {Giacca}}, \ and\ \bibinfo {author} {\bibfnamefont
  {M.}~\bibnamefont {Lusic}},\ }\href {\doibase 10.1038/nature14226} {\bibfield
   {journal} {\bibinfo  {journal} {Nature}\ }\textbf {\bibinfo {volume}
  {521}},\ \bibinfo {pages} {227} (\bibinfo {year} {2015})}\BibitemShut
  {NoStop}%
\bibitem [{\citenamefont {Sultana}\ \emph {et~al.}(2019)\citenamefont
  {Sultana}, \citenamefont {van Essen}, \citenamefont {Siol}, \citenamefont
  {Bailly-Bechet}, \citenamefont {Philippe}, \citenamefont {{Zine El
  Aabidine}}, \citenamefont {Pioger}, \citenamefont {Nigumann}, \citenamefont
  {Saccani}, \citenamefont {Andrau}, \citenamefont {Gilbert},\ and\
  \citenamefont {Cristofari}}]{Sultana2019}%
  \BibitemOpen
  \bibfield  {author} {\bibinfo {author} {\bibfnamefont {T.}~\bibnamefont
  {Sultana}}, \bibinfo {author} {\bibfnamefont {D.}~\bibnamefont {van Essen}},
  \bibinfo {author} {\bibfnamefont {O.}~\bibnamefont {Siol}}, \bibinfo {author}
  {\bibfnamefont {M.}~\bibnamefont {Bailly-Bechet}}, \bibinfo {author}
  {\bibfnamefont {C.}~\bibnamefont {Philippe}}, \bibinfo {author}
  {\bibfnamefont {A.}~\bibnamefont {{Zine El Aabidine}}}, \bibinfo {author}
  {\bibfnamefont {L.}~\bibnamefont {Pioger}}, \bibinfo {author} {\bibfnamefont
  {P.}~\bibnamefont {Nigumann}}, \bibinfo {author} {\bibfnamefont
  {S.}~\bibnamefont {Saccani}}, \bibinfo {author} {\bibfnamefont {J.~C.}\
  \bibnamefont {Andrau}}, \bibinfo {author} {\bibfnamefont {N.}~\bibnamefont
  {Gilbert}}, \ and\ \bibinfo {author} {\bibfnamefont {G.}~\bibnamefont
  {Cristofari}},\ }\href {\doibase 10.1016/j.molcel.2019.02.036} {\bibfield
  {journal} {\bibinfo  {journal} {Molecular Cell}\ }\textbf {\bibinfo {volume}
  {74}},\ \bibinfo {pages} {555} (\bibinfo {year} {2019})}\BibitemShut
  {NoStop}%
\bibitem [{\citenamefont {Schmidt}\ \emph {et~al.}(2012)\citenamefont
  {Schmidt}, \citenamefont {Schwalie}, \citenamefont {Wilson}, \citenamefont
  {Ballester}, \citenamefont {Gonalves}, \citenamefont {Kutter}, \citenamefont
  {Brown}, \citenamefont {Marshall}, \citenamefont {Flicek},\ and\
  \citenamefont {Odom}}]{Schmidt2012}%
  \BibitemOpen
  \bibfield  {author} {\bibinfo {author} {\bibfnamefont {D.}~\bibnamefont
  {Schmidt}}, \bibinfo {author} {\bibfnamefont {P.~C.}\ \bibnamefont
  {Schwalie}}, \bibinfo {author} {\bibfnamefont {M.~D.}\ \bibnamefont
  {Wilson}}, \bibinfo {author} {\bibfnamefont {B.}~\bibnamefont {Ballester}},
  \bibinfo {author} {\bibfnamefont {{\^{A}}.}~\bibnamefont {Gonalves}},
  \bibinfo {author} {\bibfnamefont {C.}~\bibnamefont {Kutter}}, \bibinfo
  {author} {\bibfnamefont {G.~D.}\ \bibnamefont {Brown}}, \bibinfo {author}
  {\bibfnamefont {A.}~\bibnamefont {Marshall}}, \bibinfo {author}
  {\bibfnamefont {P.}~\bibnamefont {Flicek}}, \ and\ \bibinfo {author}
  {\bibfnamefont {D.~T.}\ \bibnamefont {Odom}},\ }\href@noop {} {\bibfield
  {journal} {\bibinfo  {journal} {Cell}\ }\textbf {\bibinfo {volume} {148}},\
  \bibinfo {pages} {335} (\bibinfo {year} {2012})}\BibitemShut {NoStop}%
\bibitem [{\citenamefont {Sun}\ \emph {et~al.}(2018)\citenamefont {Sun},
  \citenamefont {Zhou}, \citenamefont {Emerson}, \citenamefont {Phyo},
  \citenamefont {Titus}, \citenamefont {Gong}, \citenamefont {Gilgenast},
  \citenamefont {Beagan}, \citenamefont {Davidson}, \citenamefont {Tassone},\
  and\ \citenamefont {Phillips-Cremins}}]{Sun2018}%
  \BibitemOpen
  \bibfield  {author} {\bibinfo {author} {\bibfnamefont {J.~H.}\ \bibnamefont
  {Sun}}, \bibinfo {author} {\bibfnamefont {L.}~\bibnamefont {Zhou}}, \bibinfo
  {author} {\bibfnamefont {D.~J.}\ \bibnamefont {Emerson}}, \bibinfo {author}
  {\bibfnamefont {S.~A.}\ \bibnamefont {Phyo}}, \bibinfo {author}
  {\bibfnamefont {K.~R.}\ \bibnamefont {Titus}}, \bibinfo {author}
  {\bibfnamefont {W.}~\bibnamefont {Gong}}, \bibinfo {author} {\bibfnamefont
  {T.~G.}\ \bibnamefont {Gilgenast}}, \bibinfo {author} {\bibfnamefont {J.~A.}\
  \bibnamefont {Beagan}}, \bibinfo {author} {\bibfnamefont {B.~L.}\
  \bibnamefont {Davidson}}, \bibinfo {author} {\bibfnamefont {F.}~\bibnamefont
  {Tassone}}, \ and\ \bibinfo {author} {\bibfnamefont {J.~E.}\ \bibnamefont
  {Phillips-Cremins}},\ }\href {\doibase 10.1016/j.cell.2018.08.005} {\bibfield
   {journal} {\bibinfo  {journal} {Cell}\ }\textbf {\bibinfo {volume} {175}},\
  \bibinfo {pages} {224} (\bibinfo {year} {2018})}\BibitemShut {NoStop}%
\bibitem [{\citenamefont {Grob}\ and\ \citenamefont
  {Grossniklaus}(2019)}]{Grob2019}%
  \BibitemOpen
  \bibfield  {author} {\bibinfo {author} {\bibfnamefont {S.}~\bibnamefont
  {Grob}}\ and\ \bibinfo {author} {\bibfnamefont {U.}~\bibnamefont
  {Grossniklaus}},\ }\href {\doibase 10.1186/s13059-019-1722-3} {\bibfield
  {journal} {\bibinfo  {journal} {Genome Biology}\ }\textbf {\bibinfo {volume}
  {20}},\ \bibinfo {pages} {1} (\bibinfo {year} {2019})}\BibitemShut {NoStop}%
\bibitem [{\citenamefont {Jupe}\ \emph {et~al.}(2019)\citenamefont {Jupe},
  \citenamefont {Rivkin}, \citenamefont {Michael}, \citenamefont {Zander},
  \citenamefont {Motley}, \citenamefont {Sandoval}, \citenamefont {{Keith
  Slotkin}}, \citenamefont {Chen}, \citenamefont {Castanon}, \citenamefont
  {Nery},\ and\ \citenamefont {Ecker}}]{Jupe2019}%
  \BibitemOpen
  \bibfield  {author} {\bibinfo {author} {\bibfnamefont {F.}~\bibnamefont
  {Jupe}}, \bibinfo {author} {\bibfnamefont {A.~C.}\ \bibnamefont {Rivkin}},
  \bibinfo {author} {\bibfnamefont {T.~P.}\ \bibnamefont {Michael}}, \bibinfo
  {author} {\bibfnamefont {M.}~\bibnamefont {Zander}}, \bibinfo {author}
  {\bibfnamefont {S.~T.}\ \bibnamefont {Motley}}, \bibinfo {author}
  {\bibfnamefont {J.~P.}\ \bibnamefont {Sandoval}}, \bibinfo {author}
  {\bibfnamefont {R.}~\bibnamefont {{Keith Slotkin}}}, \bibinfo {author}
  {\bibfnamefont {H.}~\bibnamefont {Chen}}, \bibinfo {author} {\bibfnamefont
  {R.}~\bibnamefont {Castanon}}, \bibinfo {author} {\bibfnamefont {J.~R.}\
  \bibnamefont {Nery}}, \ and\ \bibinfo {author} {\bibfnamefont {J.~R.}\
  \bibnamefont {Ecker}},\ }\href {\doibase 10.1371/journal.pgen.1007819}
  {\bibfield  {journal} {\bibinfo  {journal} {PLoS Genetics}\ }\textbf
  {\bibinfo {volume} {15}},\ \bibinfo {pages} {1} (\bibinfo {year}
  {2019})}\BibitemShut {NoStop}%
\bibitem [{\citenamefont {Dong}\ \emph {et~al.}(2018)\citenamefont {Dong},
  \citenamefont {Li}, \citenamefont {Li}, \citenamefont {Yuan}, \citenamefont
  {Xie}, \citenamefont {Wang}, \citenamefont {Li}, \citenamefont {Yu},
  \citenamefont {Wang}, \citenamefont {Ding}, \citenamefont {Zhang},
  \citenamefont {Li}, \citenamefont {Bian}, \citenamefont {Zhang},
  \citenamefont {Wu}, \citenamefont {Liu},\ and\ \citenamefont
  {Gong}}]{Dong2018}%
  \BibitemOpen
  \bibfield  {author} {\bibinfo {author} {\bibfnamefont {Q.}~\bibnamefont
  {Dong}}, \bibinfo {author} {\bibfnamefont {N.}~\bibnamefont {Li}}, \bibinfo
  {author} {\bibfnamefont {X.}~\bibnamefont {Li}}, \bibinfo {author}
  {\bibfnamefont {Z.}~\bibnamefont {Yuan}}, \bibinfo {author} {\bibfnamefont
  {D.}~\bibnamefont {Xie}}, \bibinfo {author} {\bibfnamefont {X.}~\bibnamefont
  {Wang}}, \bibinfo {author} {\bibfnamefont {J.}~\bibnamefont {Li}}, \bibinfo
  {author} {\bibfnamefont {Y.}~\bibnamefont {Yu}}, \bibinfo {author}
  {\bibfnamefont {J.}~\bibnamefont {Wang}}, \bibinfo {author} {\bibfnamefont
  {B.}~\bibnamefont {Ding}}, \bibinfo {author} {\bibfnamefont {Z.}~\bibnamefont
  {Zhang}}, \bibinfo {author} {\bibfnamefont {C.}~\bibnamefont {Li}}, \bibinfo
  {author} {\bibfnamefont {Y.}~\bibnamefont {Bian}}, \bibinfo {author}
  {\bibfnamefont {A.}~\bibnamefont {Zhang}}, \bibinfo {author} {\bibfnamefont
  {Y.}~\bibnamefont {Wu}}, \bibinfo {author} {\bibfnamefont {B.}~\bibnamefont
  {Liu}}, \ and\ \bibinfo {author} {\bibfnamefont {L.}~\bibnamefont {Gong}},\
  }\href {\doibase 10.1111/tpj.13925} {\bibfield  {journal} {\bibinfo
  {journal} {Plant Journal}\ }\textbf {\bibinfo {volume} {94}},\ \bibinfo
  {pages} {1141} (\bibinfo {year} {2018})}\BibitemShut {NoStop}%
\bibitem [{\citenamefont {Raviram}\ \emph {et~al.}(2018)\citenamefont
  {Raviram}, \citenamefont {Rocha}, \citenamefont {Luo}, \citenamefont
  {Swanzey}, \citenamefont {Miraldi}, \citenamefont {Chuong}, \citenamefont
  {Feschotte}, \citenamefont {Bonneau},\ and\ \citenamefont
  {Skok}}]{Raviram2018}%
  \BibitemOpen
  \bibfield  {author} {\bibinfo {author} {\bibfnamefont {R.}~\bibnamefont
  {Raviram}}, \bibinfo {author} {\bibfnamefont {P.~P.}\ \bibnamefont {Rocha}},
  \bibinfo {author} {\bibfnamefont {V.~M.}\ \bibnamefont {Luo}}, \bibinfo
  {author} {\bibfnamefont {E.}~\bibnamefont {Swanzey}}, \bibinfo {author}
  {\bibfnamefont {E.~R.}\ \bibnamefont {Miraldi}}, \bibinfo {author}
  {\bibfnamefont {E.~B.}\ \bibnamefont {Chuong}}, \bibinfo {author}
  {\bibfnamefont {C.}~\bibnamefont {Feschotte}}, \bibinfo {author}
  {\bibfnamefont {R.}~\bibnamefont {Bonneau}}, \ and\ \bibinfo {author}
  {\bibfnamefont {J.~A.}\ \bibnamefont {Skok}},\ }\href {\doibase
  10.1186/s13059-018-1598-7} {\bibfield  {journal} {\bibinfo  {journal} {Genome
  Biology}\ }\textbf {\bibinfo {volume} {28}},\ \bibinfo {pages} {1} (\bibinfo
  {year} {2018})}\BibitemShut {NoStop}%
\bibitem [{\citenamefont {Cournac}\ \emph {et~al.}(2016)\citenamefont
  {Cournac}, \citenamefont {Koszul},\ and\ \citenamefont
  {Mozziconacci}}]{Cournac2016}%
  \BibitemOpen
  \bibfield  {author} {\bibinfo {author} {\bibfnamefont {A.}~\bibnamefont
  {Cournac}}, \bibinfo {author} {\bibfnamefont {R.}~\bibnamefont {Koszul}}, \
  and\ \bibinfo {author} {\bibfnamefont {J.}~\bibnamefont {Mozziconacci}},\
  }\href {\doibase 10.1093/nar/gkv1292} {\bibfield  {journal} {\bibinfo
  {journal} {Nucleic Acids Research}\ }\textbf {\bibinfo {volume} {44}},\
  \bibinfo {pages} {245} (\bibinfo {year} {2016})}\BibitemShut {NoStop}%
\bibitem [{\citenamefont {Sun}\ \emph {et~al.}(2020)\citenamefont {Sun},
  \citenamefont {Jing}, \citenamefont {Liu}, \citenamefont {Li}, \citenamefont
  {Xue}, \citenamefont {Cheng}, \citenamefont {Wang}, \citenamefont {He},\ and\
  \citenamefont {Qian}}]{Sun2020}%
  \BibitemOpen
  \bibfield  {author} {\bibinfo {author} {\bibfnamefont {L.}~\bibnamefont
  {Sun}}, \bibinfo {author} {\bibfnamefont {Y.}~\bibnamefont {Jing}}, \bibinfo
  {author} {\bibfnamefont {X.}~\bibnamefont {Liu}}, \bibinfo {author}
  {\bibfnamefont {Q.}~\bibnamefont {Li}}, \bibinfo {author} {\bibfnamefont
  {Z.}~\bibnamefont {Xue}}, \bibinfo {author} {\bibfnamefont {Z.}~\bibnamefont
  {Cheng}}, \bibinfo {author} {\bibfnamefont {D.}~\bibnamefont {Wang}},
  \bibinfo {author} {\bibfnamefont {H.}~\bibnamefont {He}}, \ and\ \bibinfo
  {author} {\bibfnamefont {W.}~\bibnamefont {Qian}},\ }\href@noop {} {\bibfield
   {journal} {\bibinfo  {journal} {Nature Communications}\ }\textbf {\bibinfo
  {volume} {11}} (\bibinfo {year} {2020})}\BibitemShut {NoStop}%
\bibitem [{\citenamefont {Vanderlinden}\ \emph {et~al.}(2019)\citenamefont
  {Vanderlinden}, \citenamefont {Brouns}, \citenamefont {Walker}, \citenamefont
  {Kolbeck}, \citenamefont {Milles}, \citenamefont {Ott}, \citenamefont
  {Nickels}, \citenamefont {Debyser},\ and\ \citenamefont
  {Lipfert}}]{Vanderlinden2019}%
  \BibitemOpen
  \bibfield  {author} {\bibinfo {author} {\bibfnamefont {W.}~\bibnamefont
  {Vanderlinden}}, \bibinfo {author} {\bibfnamefont {T.}~\bibnamefont
  {Brouns}}, \bibinfo {author} {\bibfnamefont {P.~U.}\ \bibnamefont {Walker}},
  \bibinfo {author} {\bibfnamefont {P.~J.}\ \bibnamefont {Kolbeck}}, \bibinfo
  {author} {\bibfnamefont {L.~F.}\ \bibnamefont {Milles}}, \bibinfo {author}
  {\bibfnamefont {W.}~\bibnamefont {Ott}}, \bibinfo {author} {\bibfnamefont
  {P.~C.}\ \bibnamefont {Nickels}}, \bibinfo {author} {\bibfnamefont
  {Z.}~\bibnamefont {Debyser}}, \ and\ \bibinfo {author} {\bibfnamefont
  {J.}~\bibnamefont {Lipfert}},\ }\href@noop {} {\bibfield  {journal} {\bibinfo
   {journal} {Nature Communications}\ }\textbf {\bibinfo {volume} {10}},\
  \bibinfo {pages} {1} (\bibinfo {year} {2019})}\BibitemShut {NoStop}%
\bibitem [{\citenamefont {Prizak}\ and\ \citenamefont
  {Hilbert}(2022)}]{Prizak2022}%
  \BibitemOpen
  \bibfield  {author} {\bibinfo {author} {\bibfnamefont {R.}~\bibnamefont
  {Prizak}}\ and\ \bibinfo {author} {\bibfnamefont {L.}~\bibnamefont
  {Hilbert}},\ }\href@noop {} {\bibfield  {journal} {\bibinfo  {journal}
  {bioRxiv}\ ,\ \bibinfo {pages} {2022.02.16.480760}} (\bibinfo {year}
  {2022})}\BibitemShut {NoStop}%
\bibitem [{\citenamefont {Pasi}\ \emph {et~al.}(2016)\citenamefont {Pasi},
  \citenamefont {Mornico}, \citenamefont {Volant}, \citenamefont {Juchet},
  \citenamefont {Batisse}, \citenamefont {Bouchier}, \citenamefont {Parissi},
  \citenamefont {Ruff}, \citenamefont {Lavery},\ and\ \citenamefont
  {Lavigne}}]{Pasi2016}%
  \BibitemOpen
  \bibfield  {author} {\bibinfo {author} {\bibfnamefont {M.}~\bibnamefont
  {Pasi}}, \bibinfo {author} {\bibfnamefont {D.}~\bibnamefont {Mornico}},
  \bibinfo {author} {\bibfnamefont {S.}~\bibnamefont {Volant}}, \bibinfo
  {author} {\bibfnamefont {A.}~\bibnamefont {Juchet}}, \bibinfo {author}
  {\bibfnamefont {J.}~\bibnamefont {Batisse}}, \bibinfo {author} {\bibfnamefont
  {C.}~\bibnamefont {Bouchier}}, \bibinfo {author} {\bibfnamefont
  {V.}~\bibnamefont {Parissi}}, \bibinfo {author} {\bibfnamefont
  {M.}~\bibnamefont {Ruff}}, \bibinfo {author} {\bibfnamefont {R.}~\bibnamefont
  {Lavery}}, \ and\ \bibinfo {author} {\bibfnamefont {M.}~\bibnamefont
  {Lavigne}},\ }\href {https://academic.oup.com/nar/article/44/16/7830/2460372}
  {\bibfield  {journal} {\bibinfo  {journal} {Nucleic Acids Research}\ }\textbf
  {\bibinfo {volume} {44}},\ \bibinfo {pages} {7830} (\bibinfo {year}
  {2016})}\BibitemShut {NoStop}%
\bibitem [{\citenamefont {Michieletto}\ \emph {et~al.}(2019)\citenamefont
  {Michieletto}, \citenamefont {Lusic}, \citenamefont {Marenduzzo},\ and\
  \citenamefont {Orlandini}}]{Michieletto2019}%
  \BibitemOpen
  \bibfield  {author} {\bibinfo {author} {\bibfnamefont {D.}~\bibnamefont
  {Michieletto}}, \bibinfo {author} {\bibfnamefont {M.}~\bibnamefont {Lusic}},
  \bibinfo {author} {\bibfnamefont {D.}~\bibnamefont {Marenduzzo}}, \ and\
  \bibinfo {author} {\bibfnamefont {E.}~\bibnamefont {Orlandini}},\ }\href
  {\doibase 10.1038/s41467-019-08333-8} {\bibfield  {journal} {\bibinfo
  {journal} {Nature Communications}\ }\textbf {\bibinfo {volume} {10}},\
  \bibinfo {pages} {1} (\bibinfo {year} {2019})}\BibitemShut {NoStop}%
\bibitem [{\citenamefont {Bousios}\ \emph {et~al.}(2020)\citenamefont
  {Bousios}, \citenamefont {Nuetzmann}, \citenamefont {Buck},\ and\
  \citenamefont {Michieletto}}]{Bousios2020}%
  \BibitemOpen
  \bibfield  {author} {\bibinfo {author} {\bibfnamefont {A.}~\bibnamefont
  {Bousios}}, \bibinfo {author} {\bibfnamefont {H.-W.}\ \bibnamefont
  {Nuetzmann}}, \bibinfo {author} {\bibfnamefont {D.}~\bibnamefont {Buck}}, \
  and\ \bibinfo {author} {\bibfnamefont {D.}~\bibnamefont {Michieletto}},\
  }\href@noop {} {\bibfield  {journal} {\bibinfo  {journal} {Mobile DNA}\
  }\textbf {\bibinfo {volume} {11}} (\bibinfo {year} {2020})}\BibitemShut
  {NoStop}%
\bibitem [{\citenamefont {Pruss}\ \emph
  {et~al.}(1994{\natexlab{a}})\citenamefont {Pruss}, \citenamefont {Reeves},
  \citenamefont {Bushman},\ and\ \citenamefont {Wolffe}}]{Pruss1994}%
  \BibitemOpen
  \bibfield  {author} {\bibinfo {author} {\bibfnamefont {D.}~\bibnamefont
  {Pruss}}, \bibinfo {author} {\bibfnamefont {R.}~\bibnamefont {Reeves}},
  \bibinfo {author} {\bibfnamefont {F.}~\bibnamefont {Bushman}}, \ and\
  \bibinfo {author} {\bibfnamefont {A.}~\bibnamefont {Wolffe}},\ }\href
  {http://www.jbc.org/content/269/40/25031.short} {\bibfield  {journal}
  {\bibinfo  {journal} {J. Biol. Chem.}\ }\textbf {\bibinfo {volume} {269}},\
  \bibinfo {pages} {25031} (\bibinfo {year} {1994}{\natexlab{a}})}\BibitemShut
  {NoStop}%
\bibitem [{\citenamefont {Benleulmi}\ \emph {et~al.}(2015)\citenamefont
  {Benleulmi}, \citenamefont {Matysiak}, \citenamefont {Henriquez},
  \citenamefont {Vaillant}, \citenamefont {Lesbats}, \citenamefont {Calmels},
  \citenamefont {Naughtin}, \citenamefont {Leon}, \citenamefont {Skalka},
  \citenamefont {Ruff}, \citenamefont {Lavigne}, \citenamefont {Andreola},\
  and\ \citenamefont {Parissi}}]{Benleulmi2015}%
  \BibitemOpen
  \bibfield  {author} {\bibinfo {author} {\bibfnamefont {M.}~\bibnamefont
  {Benleulmi}}, \bibinfo {author} {\bibfnamefont {J.}~\bibnamefont {Matysiak}},
  \bibinfo {author} {\bibfnamefont {D.}~\bibnamefont {Henriquez}}, \bibinfo
  {author} {\bibfnamefont {C.}~\bibnamefont {Vaillant}}, \bibinfo {author}
  {\bibfnamefont {P.}~\bibnamefont {Lesbats}}, \bibinfo {author} {\bibfnamefont
  {C.}~\bibnamefont {Calmels}}, \bibinfo {author} {\bibfnamefont
  {M.}~\bibnamefont {Naughtin}}, \bibinfo {author} {\bibfnamefont
  {O.}~\bibnamefont {Leon}}, \bibinfo {author} {\bibfnamefont {A.}~\bibnamefont
  {Skalka}}, \bibinfo {author} {\bibfnamefont {M.}~\bibnamefont {Ruff}},
  \bibinfo {author} {\bibfnamefont {M.}~\bibnamefont {Lavigne}}, \bibinfo
  {author} {\bibfnamefont {M.-L.}\ \bibnamefont {Andreola}}, \ and\ \bibinfo
  {author} {\bibfnamefont {V.}~\bibnamefont {Parissi}},\ }\href@noop {}
  {\bibfield  {journal} {\bibinfo  {journal} {Retrovirology}\ }\textbf
  {\bibinfo {volume} {12}},\ \bibinfo {pages} {13} (\bibinfo {year}
  {2015})}\BibitemShut {NoStop}%
\bibitem [{\citenamefont {Naughtin}\ \emph {et~al.}(2015)\citenamefont
  {Naughtin}, \citenamefont {Haftek-Terreau}, \citenamefont {Xavier},
  \citenamefont {Meyer}, \citenamefont {Silvain}, \citenamefont {Jaszczyszyn},
  \citenamefont {Levy}, \citenamefont {Miele}, \citenamefont {Benleulmi},
  \citenamefont {Ruff}, \citenamefont {Parissi}, \citenamefont {Vaillant},\
  and\ \citenamefont {Lavigne}}]{Naughtin2015}%
  \BibitemOpen
  \bibfield  {author} {\bibinfo {author} {\bibfnamefont {M.}~\bibnamefont
  {Naughtin}}, \bibinfo {author} {\bibfnamefont {Z.}~\bibnamefont
  {Haftek-Terreau}}, \bibinfo {author} {\bibfnamefont {J.}~\bibnamefont
  {Xavier}}, \bibinfo {author} {\bibfnamefont {S.}~\bibnamefont {Meyer}},
  \bibinfo {author} {\bibfnamefont {M.}~\bibnamefont {Silvain}}, \bibinfo
  {author} {\bibfnamefont {Y.}~\bibnamefont {Jaszczyszyn}}, \bibinfo {author}
  {\bibfnamefont {N.}~\bibnamefont {Levy}}, \bibinfo {author} {\bibfnamefont
  {V.}~\bibnamefont {Miele}}, \bibinfo {author} {\bibfnamefont {M.~S.}\
  \bibnamefont {Benleulmi}}, \bibinfo {author} {\bibfnamefont {M.}~\bibnamefont
  {Ruff}}, \bibinfo {author} {\bibfnamefont {V.}~\bibnamefont {Parissi}},
  \bibinfo {author} {\bibfnamefont {C.}~\bibnamefont {Vaillant}}, \ and\
  \bibinfo {author} {\bibfnamefont {M.}~\bibnamefont {Lavigne}},\ }\href@noop
  {} {\bibfield  {journal} {\bibinfo  {journal} {PLoS ONE}\ }\textbf {\bibinfo
  {volume} {10}},\ \bibinfo {pages} {1} (\bibinfo {year} {2015})}\BibitemShut
  {NoStop}%
\bibitem [{\citenamefont {Matysiak}\ \emph {et~al.}(2017)\citenamefont
  {Matysiak}, \citenamefont {Lesbats}, \citenamefont {Mauro}, \citenamefont
  {Lapaillerie}, \citenamefont {Dupuy}, \citenamefont {Lopez}, \citenamefont
  {Benleulmi}, \citenamefont {Calmels}, \citenamefont {Andreola}, \citenamefont
  {Ruff}, \citenamefont {Llano}, \citenamefont {Delelis}, \citenamefont
  {Lavigne},\ and\ \citenamefont {Parissi}}]{Matysiak2017}%
  \BibitemOpen
  \bibfield  {author} {\bibinfo {author} {\bibfnamefont {J.}~\bibnamefont
  {Matysiak}}, \bibinfo {author} {\bibfnamefont {P.}~\bibnamefont {Lesbats}},
  \bibinfo {author} {\bibfnamefont {E.}~\bibnamefont {Mauro}}, \bibinfo
  {author} {\bibfnamefont {D.}~\bibnamefont {Lapaillerie}}, \bibinfo {author}
  {\bibfnamefont {J.-W.}\ \bibnamefont {Dupuy}}, \bibinfo {author}
  {\bibfnamefont {A.~P.}\ \bibnamefont {Lopez}}, \bibinfo {author}
  {\bibfnamefont {M.~S.}\ \bibnamefont {Benleulmi}}, \bibinfo {author}
  {\bibfnamefont {C.}~\bibnamefont {Calmels}}, \bibinfo {author} {\bibfnamefont
  {M.-L.}\ \bibnamefont {Andreola}}, \bibinfo {author} {\bibfnamefont
  {M.}~\bibnamefont {Ruff}}, \bibinfo {author} {\bibfnamefont {M.}~\bibnamefont
  {Llano}}, \bibinfo {author} {\bibfnamefont {O.}~\bibnamefont {Delelis}},
  \bibinfo {author} {\bibfnamefont {M.}~\bibnamefont {Lavigne}}, \ and\
  \bibinfo {author} {\bibfnamefont {V.}~\bibnamefont {Parissi}},\ }\href
  {\doibase 10.1186/s12977-017-0363-4} {\bibfield  {journal} {\bibinfo
  {journal} {Retrovirology}\ }\textbf {\bibinfo {volume} {14}},\ \bibinfo
  {pages} {39} (\bibinfo {year} {2017})}\BibitemShut {NoStop}%
\bibitem [{\citenamefont {Buenrostro}\ \emph {et~al.}(2013)\citenamefont
  {Buenrostro}, \citenamefont {Giresi}, \citenamefont {Zaba}, \citenamefont
  {Chang},\ and\ \citenamefont {Greenleaf}}]{Buenrostro2013}%
  \BibitemOpen
  \bibfield  {author} {\bibinfo {author} {\bibfnamefont {J.~D.}\ \bibnamefont
  {Buenrostro}}, \bibinfo {author} {\bibfnamefont {P.~G.}\ \bibnamefont
  {Giresi}}, \bibinfo {author} {\bibfnamefont {L.~C.}\ \bibnamefont {Zaba}},
  \bibinfo {author} {\bibfnamefont {H.~Y.}\ \bibnamefont {Chang}}, \ and\
  \bibinfo {author} {\bibfnamefont {W.~J.}\ \bibnamefont {Greenleaf}},\ }\href
  {\doibase 10.1038/nmeth.2688} {\bibfield  {journal} {\bibinfo  {journal}
  {Nature Methods}\ }\textbf {\bibinfo {volume} {10}},\ \bibinfo {pages} {1213}
  (\bibinfo {year} {2013})}\BibitemShut {NoStop}%
\bibitem [{\citenamefont {Maeshima}\ \emph {et~al.}(2015)\citenamefont
  {Maeshima}, \citenamefont {Kaizu}, \citenamefont {Tamura}, \citenamefont
  {Nozaki}, \citenamefont {Kokubo},\ and\ \citenamefont
  {Takahashi}}]{Maeshima2015}%
  \BibitemOpen
  \bibfield  {author} {\bibinfo {author} {\bibfnamefont {K.}~\bibnamefont
  {Maeshima}}, \bibinfo {author} {\bibfnamefont {K.}~\bibnamefont {Kaizu}},
  \bibinfo {author} {\bibfnamefont {S.}~\bibnamefont {Tamura}}, \bibinfo
  {author} {\bibfnamefont {T.}~\bibnamefont {Nozaki}}, \bibinfo {author}
  {\bibfnamefont {T.}~\bibnamefont {Kokubo}}, \ and\ \bibinfo {author}
  {\bibfnamefont {K.}~\bibnamefont {Takahashi}},\ }\href@noop {} {\bibfield
  {journal} {\bibinfo  {journal} {Journal of Physics Condensed Matter}\
  }\textbf {\bibinfo {volume} {27}} (\bibinfo {year} {2015})}\BibitemShut
  {NoStop}%
\bibitem [{\citenamefont {Shin}\ \emph {et~al.}(2018)\citenamefont {Shin},
  \citenamefont {Chang}, \citenamefont {Lee}, \citenamefont {Berry},
  \citenamefont {Sanders}, \citenamefont {Ronceray}, \citenamefont {Wingreen},
  \citenamefont {Haataja},\ and\ \citenamefont {Brangwynne}}]{Shin2018a}%
  \BibitemOpen
  \bibfield  {author} {\bibinfo {author} {\bibfnamefont {Y.}~\bibnamefont
  {Shin}}, \bibinfo {author} {\bibfnamefont {Y.~C.}\ \bibnamefont {Chang}},
  \bibinfo {author} {\bibfnamefont {D.~S.}\ \bibnamefont {Lee}}, \bibinfo
  {author} {\bibfnamefont {J.}~\bibnamefont {Berry}}, \bibinfo {author}
  {\bibfnamefont {D.~W.}\ \bibnamefont {Sanders}}, \bibinfo {author}
  {\bibfnamefont {P.}~\bibnamefont {Ronceray}}, \bibinfo {author}
  {\bibfnamefont {N.~S.}\ \bibnamefont {Wingreen}}, \bibinfo {author}
  {\bibfnamefont {M.}~\bibnamefont {Haataja}}, \ and\ \bibinfo {author}
  {\bibfnamefont {C.~P.}\ \bibnamefont {Brangwynne}},\ }\href {\doibase
  10.1016/j.cell.2018.10.057} {\bibfield  {journal} {\bibinfo  {journal}
  {Cell}\ }\textbf {\bibinfo {volume} {175}},\ \bibinfo {pages} {1481}
  (\bibinfo {year} {2018})}\BibitemShut {NoStop}%
\bibitem [{\citenamefont {Bosse}\ \emph {et~al.}(2015)\citenamefont {Bosse},
  \citenamefont {Hogue}, \citenamefont {Feric}, \citenamefont {Thiberge},
  \citenamefont {Sodeik}, \citenamefont {Brangwynne},\ and\ \citenamefont
  {Enquist}}]{Bosse2015}%
  \BibitemOpen
  \bibfield  {author} {\bibinfo {author} {\bibfnamefont {J.~B.}\ \bibnamefont
  {Bosse}}, \bibinfo {author} {\bibfnamefont {I.~B.}\ \bibnamefont {Hogue}},
  \bibinfo {author} {\bibfnamefont {M.}~\bibnamefont {Feric}}, \bibinfo
  {author} {\bibfnamefont {S.~Y.}\ \bibnamefont {Thiberge}}, \bibinfo {author}
  {\bibfnamefont {B.}~\bibnamefont {Sodeik}}, \bibinfo {author} {\bibfnamefont
  {C.~P.}\ \bibnamefont {Brangwynne}}, \ and\ \bibinfo {author} {\bibfnamefont
  {L.~W.}\ \bibnamefont {Enquist}},\ }\href {\doibase 10.1073/pnas.1513876112}
  {\emph {\bibinfo {title} {Proc. Nat. Acad. Sci. USA}}},\ Vol.\ \bibinfo
  {volume} {112}\ (\bibinfo {year} {2015})\ pp.\ \bibinfo {pages}
  {E5725--E5733}\BibitemShut {NoStop}%
\bibitem [{\citenamefont {Burdick}\ \emph {et~al.}(2020)\citenamefont
  {Burdick}, \citenamefont {Li}, \citenamefont {Munshi}, \citenamefont
  {Rawson}, \citenamefont {Nagashima}, \citenamefont {Hu},\ and\ \citenamefont
  {Pathak}}]{Burdick2020}%
  \BibitemOpen
  \bibfield  {author} {\bibinfo {author} {\bibfnamefont {R.~C.}\ \bibnamefont
  {Burdick}}, \bibinfo {author} {\bibfnamefont {C.}~\bibnamefont {Li}},
  \bibinfo {author} {\bibfnamefont {M.~H.}\ \bibnamefont {Munshi}}, \bibinfo
  {author} {\bibfnamefont {J.~M.}\ \bibnamefont {Rawson}}, \bibinfo {author}
  {\bibfnamefont {K.}~\bibnamefont {Nagashima}}, \bibinfo {author}
  {\bibfnamefont {W.~S.}\ \bibnamefont {Hu}}, \ and\ \bibinfo {author}
  {\bibfnamefont {V.~K.}\ \bibnamefont {Pathak}},\ }\href {\doibase
  10.1073/pnas.1920631117} {\bibfield  {journal} {\bibinfo  {journal}
  {Proceedings of the National Academy of Sciences of the United States of
  America}\ }\textbf {\bibinfo {volume} {117}},\ \bibinfo {pages} {5486}
  (\bibinfo {year} {2020})}\BibitemShut {NoStop}%
\bibitem [{\citenamefont {Pruss}\ \emph
  {et~al.}(1994{\natexlab{b}})\citenamefont {Pruss}, \citenamefont {Bushman},\
  and\ \citenamefont {Wolffe}}]{Pruss1994a}%
  \BibitemOpen
  \bibfield  {author} {\bibinfo {author} {\bibfnamefont {D.}~\bibnamefont
  {Pruss}}, \bibinfo {author} {\bibfnamefont {F.}~\bibnamefont {Bushman}}, \
  and\ \bibinfo {author} {\bibfnamefont {A.}~\bibnamefont {Wolffe}},\ }\href
  {http://www.pnas.org/content/91/13/5913.short} {\bibfield  {journal}
  {\bibinfo  {journal} {Proc. Natl. Acad. Sci. USA}\ }\textbf {\bibinfo
  {volume} {91}},\ \bibinfo {pages} {5913} (\bibinfo {year}
  {1994}{\natexlab{b}})}\BibitemShut {NoStop}%
\bibitem [{\citenamefont {Jones}\ \emph {et~al.}(2016)\citenamefont {Jones},
  \citenamefont {Lopez}, \citenamefont {Hanne}, \citenamefont {Peake},
  \citenamefont {Lee}, \citenamefont {Fishel},\ and\ \citenamefont
  {Yoder}}]{Jones2016a}%
  \BibitemOpen
  \bibfield  {author} {\bibinfo {author} {\bibfnamefont {N.~D.}\ \bibnamefont
  {Jones}}, \bibinfo {author} {\bibfnamefont {M.~A.}\ \bibnamefont {Lopez}},
  \bibinfo {author} {\bibfnamefont {J.}~\bibnamefont {Hanne}}, \bibinfo
  {author} {\bibfnamefont {M.~B.}\ \bibnamefont {Peake}}, \bibinfo {author}
  {\bibfnamefont {J.~B.}\ \bibnamefont {Lee}}, \bibinfo {author} {\bibfnamefont
  {R.}~\bibnamefont {Fishel}}, \ and\ \bibinfo {author} {\bibfnamefont {K.~E.}\
  \bibnamefont {Yoder}},\ }\href {\doibase 10.1038/ncomms11409} {\bibfield
  {journal} {\bibinfo  {journal} {Nature Communications}\ }\textbf {\bibinfo
  {volume} {7}},\ \bibinfo {pages} {1} (\bibinfo {year} {2016})}\BibitemShut
  {NoStop}%
\bibitem [{\citenamefont {Kvaratskhelia}\ \emph {et~al.}(2014)\citenamefont
  {Kvaratskhelia}, \citenamefont {Sharma}, \citenamefont {Larue}, \citenamefont
  {Serrao},\ and\ \citenamefont {Engelman}}]{Kvaratskhelia2014}%
  \BibitemOpen
  \bibfield  {author} {\bibinfo {author} {\bibfnamefont {M.}~\bibnamefont
  {Kvaratskhelia}}, \bibinfo {author} {\bibfnamefont {A.}~\bibnamefont
  {Sharma}}, \bibinfo {author} {\bibfnamefont {R.~C.}\ \bibnamefont {Larue}},
  \bibinfo {author} {\bibfnamefont {E.}~\bibnamefont {Serrao}}, \ and\ \bibinfo
  {author} {\bibfnamefont {A.}~\bibnamefont {Engelman}},\ }\href {\doibase
  10.1093/nar/gku769} {\bibfield  {journal} {\bibinfo  {journal} {Nucleic Acids
  Res.}\ }\textbf {\bibinfo {volume} {42}},\ \bibinfo {pages} {10209} (\bibinfo
  {year} {2014})}\BibitemShut {NoStop}%
\bibitem [{\citenamefont {Frenkel}\ and\ \citenamefont
  {Smit}(1996)}]{Frenkel2001}%
  \BibitemOpen
  \bibfield  {author} {\bibinfo {author} {\bibfnamefont {D.}~\bibnamefont
  {Frenkel}}\ and\ \bibinfo {author} {\bibfnamefont {B.}~\bibnamefont {Smit}},\
  }\href@noop {} {\emph {\bibinfo {title} {Understanding molecular simulation:
  From algorithms to applications}}}\ (\bibinfo {year} {1996})\BibitemShut
  {NoStop}%
\bibitem [{\citenamefont {Plimpton}(1995)}]{Plimpton1995}%
  \BibitemOpen
  \bibfield  {author} {\bibinfo {author} {\bibfnamefont {S.}~\bibnamefont
  {Plimpton}},\ }\href@noop {} {\bibfield  {journal} {\bibinfo  {journal} {J.
  Comp. Phys.}\ }\textbf {\bibinfo {volume} {117}},\ \bibinfo {pages} {1}
  (\bibinfo {year} {1995})}\BibitemShut {NoStop}%
\bibitem [{\citenamefont {Sides}\ \emph {et~al.}(2004)\citenamefont {Sides},
  \citenamefont {Grest}, \citenamefont {Stevens},\ and\ \citenamefont
  {Plimpton}}]{Sides2004}%
  \BibitemOpen
  \bibfield  {author} {\bibinfo {author} {\bibfnamefont {S.~W.}\ \bibnamefont
  {Sides}}, \bibinfo {author} {\bibfnamefont {G.~S.}\ \bibnamefont {Grest}},
  \bibinfo {author} {\bibfnamefont {M.~J.}\ \bibnamefont {Stevens}}, \ and\
  \bibinfo {author} {\bibfnamefont {S.~J.}\ \bibnamefont {Plimpton}},\
  }\href@noop {} {\bibfield  {journal} {\bibinfo  {journal} {Journal of Polymer
  Science, Part B: Polymer Physics}\ }\textbf {\bibinfo {volume} {42}},\
  \bibinfo {pages} {199} (\bibinfo {year} {2004})}\BibitemShut {NoStop}%
\bibitem [{\citenamefont {Forte}\ \emph {et~al.}(2021)\citenamefont {Forte},
  \citenamefont {Michieletto}, \citenamefont {Marenduzzo},\ and\ \citenamefont
  {Orlandini}}]{forte2021investigating}%
  \BibitemOpen
  \bibfield  {author} {\bibinfo {author} {\bibfnamefont {G.}~\bibnamefont
  {Forte}}, \bibinfo {author} {\bibfnamefont {D.}~\bibnamefont {Michieletto}},
  \bibinfo {author} {\bibfnamefont {D.}~\bibnamefont {Marenduzzo}}, \ and\
  \bibinfo {author} {\bibfnamefont {E.}~\bibnamefont {Orlandini}},\ }\href@noop
  {} {\bibfield  {journal} {\bibinfo  {journal} {Journal of the Royal Society
  Interface}\ }\textbf {\bibinfo {volume} {18}},\ \bibinfo {pages} {20210229}
  (\bibinfo {year} {2021})}\BibitemShut {NoStop}%
\bibitem [{\citenamefont {Bonato}\ \emph {et~al.}(2022)\citenamefont {Bonato},
  \citenamefont {Marenduzzo}, \citenamefont {Michieletto},\ and\ \citenamefont
  {Orlandini}}]{Bonato2022}%
  \BibitemOpen
  \bibfield  {author} {\bibinfo {author} {\bibfnamefont {A.}~\bibnamefont
  {Bonato}}, \bibinfo {author} {\bibfnamefont {D.}~\bibnamefont {Marenduzzo}},
  \bibinfo {author} {\bibfnamefont {D.}~\bibnamefont {Michieletto}}, \ and\
  \bibinfo {author} {\bibfnamefont {E.}~\bibnamefont {Orlandini}},\ }\href
  {\doibase 10.1073/pnas.2207728119} {\bibfield  {journal} {\bibinfo  {journal}
  {Proceedings of the National Academy of Sciences of the United States of
  America}\ }\textbf {\bibinfo {volume} {119}},\ \bibinfo {pages} {1} (\bibinfo
  {year} {2022})}\BibitemShut {NoStop}%
\bibitem [{\citenamefont {Kent}\ \emph {et~al.}(2002)\citenamefont {Kent},
  \citenamefont {Sugnet}, \citenamefont {Furey}, \citenamefont {Roskin},
  \citenamefont {Pringle}, \citenamefont {Zahler}, \citenamefont {Haussler},\
  and\ \citenamefont {David}}]{Kent2002}%
  \BibitemOpen
  \bibfield  {author} {\bibinfo {author} {\bibfnamefont {W.~J.}\ \bibnamefont
  {Kent}}, \bibinfo {author} {\bibfnamefont {C.~W.}\ \bibnamefont {Sugnet}},
  \bibinfo {author} {\bibfnamefont {T.~S.}\ \bibnamefont {Furey}}, \bibinfo
  {author} {\bibfnamefont {K.~M.}\ \bibnamefont {Roskin}}, \bibinfo {author}
  {\bibfnamefont {T.~H.}\ \bibnamefont {Pringle}}, \bibinfo {author}
  {\bibfnamefont {A.~M.}\ \bibnamefont {Zahler}}, \bibinfo {author}
  {\bibnamefont {Haussler}}, \ and\ \bibinfo {author} {\bibnamefont {David}},\
  }\href {\doibase 10.1101/gr.229102} {\bibfield  {journal} {\bibinfo
  {journal} {Genome Research}\ }\textbf {\bibinfo {volume} {12}},\ \bibinfo
  {pages} {996} (\bibinfo {year} {2002})}\BibitemShut {NoStop}%
\bibitem [{\citenamefont {Finzi}\ and\ \citenamefont
  {Gelles}(1995)}]{Finzi1995}%
  \BibitemOpen
  \bibfield  {author} {\bibinfo {author} {\bibfnamefont {L.}~\bibnamefont
  {Finzi}}\ and\ \bibinfo {author} {\bibfnamefont {J.}~\bibnamefont {Gelles}},\
  }\href {\doibase 10.1126/science.7824935} {\bibfield  {journal} {\bibinfo
  {journal} {Science}\ }\textbf {\bibinfo {volume} {267}},\ \bibinfo {pages}
  {378} (\bibinfo {year} {1995})}\BibitemShut {NoStop}%
\bibitem [{\citenamefont {Priest}\ \emph {et~al.}(2014)\citenamefont {Priest},
  \citenamefont {Kumar}, \citenamefont {Yan}, \citenamefont {Dunlap},
  \citenamefont {Dodd},\ and\ \citenamefont {Shearwin}}]{Priest2014}%
  \BibitemOpen
  \bibfield  {author} {\bibinfo {author} {\bibfnamefont {D.~G.}\ \bibnamefont
  {Priest}}, \bibinfo {author} {\bibfnamefont {S.}~\bibnamefont {Kumar}},
  \bibinfo {author} {\bibfnamefont {Y.}~\bibnamefont {Yan}}, \bibinfo {author}
  {\bibfnamefont {D.~D.}\ \bibnamefont {Dunlap}}, \bibinfo {author}
  {\bibfnamefont {I.~B.}\ \bibnamefont {Dodd}}, \ and\ \bibinfo {author}
  {\bibfnamefont {K.~E.}\ \bibnamefont {Shearwin}},\ }\href {\doibase
  10.1073/pnas.1410764111} {\bibfield  {journal} {\bibinfo  {journal}
  {Proceedings of the National Academy of Sciences of the United States of
  America}\ }\textbf {\bibinfo {volume} {111}},\ \bibinfo {pages} {E4449}
  (\bibinfo {year} {2014})}\BibitemShut {NoStop}%
\bibitem [{\citenamefont {Ding}\ \emph {et~al.}(2014)\citenamefont {Ding},
  \citenamefont {Manzo}, \citenamefont {Fulcrand}, \citenamefont {Leng},
  \citenamefont {Dunlap},\ and\ \citenamefont {Finzi}}]{Ding2014}%
  \BibitemOpen
  \bibfield  {author} {\bibinfo {author} {\bibfnamefont {Y.}~\bibnamefont
  {Ding}}, \bibinfo {author} {\bibfnamefont {C.}~\bibnamefont {Manzo}},
  \bibinfo {author} {\bibfnamefont {G.}~\bibnamefont {Fulcrand}}, \bibinfo
  {author} {\bibfnamefont {F.}~\bibnamefont {Leng}}, \bibinfo {author}
  {\bibfnamefont {D.}~\bibnamefont {Dunlap}}, \ and\ \bibinfo {author}
  {\bibfnamefont {L.}~\bibnamefont {Finzi}},\ }\href {\doibase
  10.1073/pnas.1320644111} {\bibfield  {journal} {\bibinfo  {journal} {Proc.
  Natl. Acad. Sci. USA}\ }\textbf {\bibinfo {volume} {111}},\ \bibinfo {pages}
  {15402} (\bibinfo {year} {2014})}\BibitemShut {NoStop}%
\bibitem [{\citenamefont {Rao}\ \emph {et~al.}(2014)\citenamefont {Rao},
  \citenamefont {Huntley}, \citenamefont {Durand}, \citenamefont {Stamenova},
  \citenamefont {Bochkov}, \citenamefont {Robinson}, \citenamefont {Sanborn},
  \citenamefont {Machol}, \citenamefont {Omer}, \citenamefont {Lander},\ and\
  \citenamefont {Aiden}}]{Rao2014}%
  \BibitemOpen
  \bibfield  {author} {\bibinfo {author} {\bibfnamefont {S.~S.~P.}\
  \bibnamefont {Rao}}, \bibinfo {author} {\bibfnamefont {M.~H.}\ \bibnamefont
  {Huntley}}, \bibinfo {author} {\bibfnamefont {N.~C.}\ \bibnamefont {Durand}},
  \bibinfo {author} {\bibfnamefont {E.~K.}\ \bibnamefont {Stamenova}}, \bibinfo
  {author} {\bibfnamefont {I.~D.}\ \bibnamefont {Bochkov}}, \bibinfo {author}
  {\bibfnamefont {J.~T.}\ \bibnamefont {Robinson}}, \bibinfo {author}
  {\bibfnamefont {A.~L.}\ \bibnamefont {Sanborn}}, \bibinfo {author}
  {\bibfnamefont {I.}~\bibnamefont {Machol}}, \bibinfo {author} {\bibfnamefont
  {A.~D.}\ \bibnamefont {Omer}}, \bibinfo {author} {\bibfnamefont {E.~S.}\
  \bibnamefont {Lander}}, \ and\ \bibinfo {author} {\bibfnamefont {E.~L.}\
  \bibnamefont {Aiden}},\ }\href {\doibase 10.1016/j.cell.2014.11.021}
  {\bibfield  {journal} {\bibinfo  {journal} {Cell}\ }\textbf {\bibinfo
  {volume} {159}},\ \bibinfo {pages} {1665} (\bibinfo {year} {2014})},\ \Eprint
  {http://arxiv.org/abs/1206.5533} {arXiv:1206.5533} \BibitemShut {NoStop}%
\bibitem [{\citenamefont {Yan}\ \emph {et~al.}(2018)\citenamefont {Yan},
  \citenamefont {Leng}, \citenamefont {Finzi},\ and\ \citenamefont
  {Dunlap}}]{Yan2018}%
  \BibitemOpen
  \bibfield  {author} {\bibinfo {author} {\bibfnamefont {Y.}~\bibnamefont
  {Yan}}, \bibinfo {author} {\bibfnamefont {F.}~\bibnamefont {Leng}}, \bibinfo
  {author} {\bibfnamefont {L.}~\bibnamefont {Finzi}}, \ and\ \bibinfo {author}
  {\bibfnamefont {D.}~\bibnamefont {Dunlap}},\ }\href {\doibase
  10.1093/nar/gky021} {\bibfield  {journal} {\bibinfo  {journal} {Nucleic Acids
  Research}\ }\textbf {\bibinfo {volume} {46}},\ \bibinfo {pages} {2370}
  (\bibinfo {year} {2018})}\BibitemShut {NoStop}%
\bibitem [{\citenamefont {Alberts}\ \emph {et~al.}(2014)\citenamefont
  {Alberts}, \citenamefont {Johnson}, \citenamefont {Lewis}, \citenamefont
  {Morgan},\ and\ \citenamefont {Raff}}]{Alberts2014}%
  \BibitemOpen
  \bibfield  {author} {\bibinfo {author} {\bibfnamefont {B.}~\bibnamefont
  {Alberts}}, \bibinfo {author} {\bibfnamefont {A.}~\bibnamefont {Johnson}},
  \bibinfo {author} {\bibfnamefont {J.}~\bibnamefont {Lewis}}, \bibinfo
  {author} {\bibfnamefont {D.}~\bibnamefont {Morgan}}, \ and\ \bibinfo {author}
  {\bibfnamefont {M.}~\bibnamefont {Raff}},\ }\href@noop {} {\emph {\bibinfo
  {title} {{Molecular Biology of the Cell}}}}\ (\bibinfo  {publisher} {Taylor
  {\&} Francis},\ \bibinfo {year} {2014})\ p.\ \bibinfo {pages}
  {1464}\BibitemShut {NoStop}%
\bibitem [{\citenamefont {Ryu}\ \emph {et~al.}(2021)\citenamefont {Ryu},
  \citenamefont {Bouchoux}, \citenamefont {Liu}, \citenamefont {Kim},
  \citenamefont {Minamino}, \citenamefont {de~Groot}, \citenamefont {Katan},
  \citenamefont {Bonato}, \citenamefont {Marenduzzo}, \citenamefont
  {Michieletto}, \citenamefont {Uhlmann},\ and\ \citenamefont
  {Dekker}}]{Ryu2021}%
  \BibitemOpen
  \bibfield  {author} {\bibinfo {author} {\bibfnamefont {J.-K.}\ \bibnamefont
  {Ryu}}, \bibinfo {author} {\bibfnamefont {C.}~\bibnamefont {Bouchoux}},
  \bibinfo {author} {\bibfnamefont {H.~W.}\ \bibnamefont {Liu}}, \bibinfo
  {author} {\bibfnamefont {E.}~\bibnamefont {Kim}}, \bibinfo {author}
  {\bibfnamefont {M.}~\bibnamefont {Minamino}}, \bibinfo {author}
  {\bibfnamefont {R.}~\bibnamefont {de~Groot}}, \bibinfo {author}
  {\bibfnamefont {A.~J.}\ \bibnamefont {Katan}}, \bibinfo {author}
  {\bibfnamefont {A.}~\bibnamefont {Bonato}}, \bibinfo {author} {\bibfnamefont
  {D.}~\bibnamefont {Marenduzzo}}, \bibinfo {author} {\bibfnamefont
  {D.}~\bibnamefont {Michieletto}}, \bibinfo {author} {\bibfnamefont
  {F.}~\bibnamefont {Uhlmann}}, \ and\ \bibinfo {author} {\bibfnamefont
  {C.}~\bibnamefont {Dekker}},\ }\href {\doibase 10.1126/sciadv.abe5905}
  {\bibfield  {journal} {\bibinfo  {journal} {Science Advances}\ }\textbf
  {\bibinfo {volume} {7}},\ \bibinfo {pages} {eabe5905} (\bibinfo {year}
  {2021})}\BibitemShut {NoStop}%
\bibitem [{\citenamefont {Davidson}\ \emph {et~al.}(2019)\citenamefont
  {Davidson}, \citenamefont {Bauer}, \citenamefont {Goetz}, \citenamefont
  {Tang}, \citenamefont {Wutz},\ and\ \citenamefont {Peters}}]{Davidson2019}%
  \BibitemOpen
  \bibfield  {author} {\bibinfo {author} {\bibfnamefont {I.~F.}\ \bibnamefont
  {Davidson}}, \bibinfo {author} {\bibfnamefont {B.}~\bibnamefont {Bauer}},
  \bibinfo {author} {\bibfnamefont {D.}~\bibnamefont {Goetz}}, \bibinfo
  {author} {\bibfnamefont {W.}~\bibnamefont {Tang}}, \bibinfo {author}
  {\bibfnamefont {G.}~\bibnamefont {Wutz}}, \ and\ \bibinfo {author}
  {\bibfnamefont {J.~M.}\ \bibnamefont {Peters}},\ }\href {\doibase
  10.1126/science.aaz3418} {\bibfield  {journal} {\bibinfo  {journal}
  {Science}\ }\textbf {\bibinfo {volume} {366}},\ \bibinfo {pages} {1338}
  (\bibinfo {year} {2019})}\BibitemShut {NoStop}%
\bibitem [{\citenamefont {Hirano}(2012)}]{Hirano2012}%
  \BibitemOpen
  \bibfield  {author} {\bibinfo {author} {\bibfnamefont {T.}~\bibnamefont
  {Hirano}},\ }\href {\doibase 10.1101/gad.194746.112} {\bibfield  {journal}
  {\bibinfo  {journal} {Genes and Development}\ }\textbf {\bibinfo {volume}
  {26}},\ \bibinfo {pages} {1659} (\bibinfo {year} {2012})}\BibitemShut
  {NoStop}%
\bibitem [{\citenamefont {Ganji}\ \emph {et~al.}(2018)\citenamefont {Ganji},
  \citenamefont {Shaltiel}, \citenamefont {Bisht}, \citenamefont {Kim},
  \citenamefont {Kalichava}, \citenamefont {Haering},\ and\ \citenamefont
  {Dekker}}]{Ganji2018}%
  \BibitemOpen
  \bibfield  {author} {\bibinfo {author} {\bibfnamefont {M.}~\bibnamefont
  {Ganji}}, \bibinfo {author} {\bibfnamefont {I.~A.}\ \bibnamefont {Shaltiel}},
  \bibinfo {author} {\bibfnamefont {S.}~\bibnamefont {Bisht}}, \bibinfo
  {author} {\bibfnamefont {E.}~\bibnamefont {Kim}}, \bibinfo {author}
  {\bibfnamefont {A.}~\bibnamefont {Kalichava}}, \bibinfo {author}
  {\bibfnamefont {C.~H.}\ \bibnamefont {Haering}}, \ and\ \bibinfo {author}
  {\bibfnamefont {C.}~\bibnamefont {Dekker}},\ }\href@noop {} {\bibfield
  {journal} {\bibinfo  {journal} {Science}\ }\textbf {\bibinfo {volume}
  {360}},\ \bibinfo {pages} {102} (\bibinfo {year} {2018})}\BibitemShut
  {NoStop}%
\bibitem [{\citenamefont {Brand{\~{a}}o}\ \emph {et~al.}(2019)\citenamefont
  {Brand{\~{a}}o}, \citenamefont {Paul}, \citenamefont {van~den Berg},
  \citenamefont {Rudner}, \citenamefont {Wang},\ and\ \citenamefont
  {Mirny}}]{Brandao2019a}%
  \BibitemOpen
  \bibfield  {author} {\bibinfo {author} {\bibfnamefont {H.~B.}\ \bibnamefont
  {Brand{\~{a}}o}}, \bibinfo {author} {\bibfnamefont {P.}~\bibnamefont {Paul}},
  \bibinfo {author} {\bibfnamefont {A.~A.}\ \bibnamefont {van~den Berg}},
  \bibinfo {author} {\bibfnamefont {D.~Z.}\ \bibnamefont {Rudner}}, \bibinfo
  {author} {\bibfnamefont {X.}~\bibnamefont {Wang}}, \ and\ \bibinfo {author}
  {\bibfnamefont {L.~A.}\ \bibnamefont {Mirny}},\ }\href {\doibase
  10.1073/pnas.1907009116} {\bibfield  {journal} {\bibinfo  {journal}
  {Proceedings of the National Academy of Sciences of the United States of
  America}\ }\textbf {\bibinfo {volume} {116}},\ \bibinfo {pages} {20489}
  (\bibinfo {year} {2019})}\BibitemShut {NoStop}%
\bibitem [{\citenamefont {Brand{\~{a}}o}\ \emph {et~al.}(2021)\citenamefont
  {Brand{\~{a}}o}, \citenamefont {Ren}, \citenamefont {Karaboja}, \citenamefont
  {Mirny},\ and\ \citenamefont {Wang}}]{Brandao2021}%
  \BibitemOpen
  \bibfield  {author} {\bibinfo {author} {\bibfnamefont {H.~B.}\ \bibnamefont
  {Brand{\~{a}}o}}, \bibinfo {author} {\bibfnamefont {Z.}~\bibnamefont {Ren}},
  \bibinfo {author} {\bibfnamefont {X.}~\bibnamefont {Karaboja}}, \bibinfo
  {author} {\bibfnamefont {L.~A.}\ \bibnamefont {Mirny}}, \ and\ \bibinfo
  {author} {\bibfnamefont {X.}~\bibnamefont {Wang}},\ }\href@noop {} {\bibfield
   {journal} {\bibinfo  {journal} {Nature Structural and Molecular Biology}\
  }\textbf {\bibinfo {volume} {28}},\ \bibinfo {pages} {642} (\bibinfo {year}
  {2021})}\BibitemShut {NoStop}%
\bibitem [{\citenamefont {Yoshua}\ \emph {et~al.}(2021)\citenamefont {Yoshua},
  \citenamefont {Watson}, \citenamefont {Howard}, \citenamefont
  {Velasco-Berrelleza}, \citenamefont {Leake},\ and\ \citenamefont
  {Noy}}]{Yoshua2021}%
  \BibitemOpen
  \bibfield  {author} {\bibinfo {author} {\bibfnamefont {S.~B.}\ \bibnamefont
  {Yoshua}}, \bibinfo {author} {\bibfnamefont {G.~D.}\ \bibnamefont {Watson}},
  \bibinfo {author} {\bibfnamefont {J.~A.~L.}\ \bibnamefont {Howard}}, \bibinfo
  {author} {\bibfnamefont {V.}~\bibnamefont {Velasco-Berrelleza}}, \bibinfo
  {author} {\bibfnamefont {M.~C.}\ \bibnamefont {Leake}}, \ and\ \bibinfo
  {author} {\bibfnamefont {A.}~\bibnamefont {Noy}},\ }\href {\doibase
  10.1093/nar/gkab641} {\bibfield  {journal} {\bibinfo  {journal} {Nucleic
  Acids Research}\ }\textbf {\bibinfo {volume} {49}},\ \bibinfo {pages} {8684}
  (\bibinfo {year} {2021})}\BibitemShut {NoStop}%
\bibitem [{\citenamefont {Le}\ \emph {et~al.}(2013)\citenamefont {Le},
  \citenamefont {Imakaev}, \citenamefont {Mirny},\ and\ \citenamefont
  {Laub}}]{Le2013}%
  \BibitemOpen
  \bibfield  {author} {\bibinfo {author} {\bibfnamefont {T.~B.}\ \bibnamefont
  {Le}}, \bibinfo {author} {\bibfnamefont {M.~V.}\ \bibnamefont {Imakaev}},
  \bibinfo {author} {\bibfnamefont {L.~A.}\ \bibnamefont {Mirny}}, \ and\
  \bibinfo {author} {\bibfnamefont {M.~T.}\ \bibnamefont {Laub}},\ }\href@noop
  {} {\bibfield  {journal} {\bibinfo  {journal} {Science}\ }\textbf {\bibinfo
  {volume} {342}},\ \bibinfo {pages} {731} (\bibinfo {year}
  {2013})}\BibitemShut {NoStop}%
\bibitem [{\citenamefont {Wilson}\ \emph {et~al.}(2019)\citenamefont {Wilson},
  \citenamefont {Renault}, \citenamefont {Maskell}, \citenamefont {Ghoneim},
  \citenamefont {Pye}, \citenamefont {Nans}, \citenamefont {Rueda},
  \citenamefont {Cherepanov},\ and\ \citenamefont {Costa}}]{Wilson2019}%
  \BibitemOpen
  \bibfield  {author} {\bibinfo {author} {\bibfnamefont {M.~D.}\ \bibnamefont
  {Wilson}}, \bibinfo {author} {\bibfnamefont {L.}~\bibnamefont {Renault}},
  \bibinfo {author} {\bibfnamefont {D.~P.}\ \bibnamefont {Maskell}}, \bibinfo
  {author} {\bibfnamefont {M.}~\bibnamefont {Ghoneim}}, \bibinfo {author}
  {\bibfnamefont {V.~E.}\ \bibnamefont {Pye}}, \bibinfo {author} {\bibfnamefont
  {A.}~\bibnamefont {Nans}}, \bibinfo {author} {\bibfnamefont {D.~S.}\
  \bibnamefont {Rueda}}, \bibinfo {author} {\bibfnamefont {P.}~\bibnamefont
  {Cherepanov}}, \ and\ \bibinfo {author} {\bibfnamefont {A.}~\bibnamefont
  {Costa}},\ }\href@noop {} {\bibfield  {journal} {\bibinfo  {journal} {Nature
  Communications}\ }\textbf {\bibinfo {volume} {10}},\ \bibinfo {pages} {1}
  (\bibinfo {year} {2019})}\BibitemShut {NoStop}%
\bibitem [{\citenamefont {Brouns}\ \emph {et~al.}(2018)\citenamefont {Brouns},
  \citenamefont {{De Keersmaecker}}, \citenamefont {Konrad}, \citenamefont
  {Kodera}, \citenamefont {Ando}, \citenamefont {Lipfert}, \citenamefont {{De
  Feyter}},\ and\ \citenamefont {Vanderlinden}}]{Brouns2018}%
  \BibitemOpen
  \bibfield  {author} {\bibinfo {author} {\bibfnamefont {T.}~\bibnamefont
  {Brouns}}, \bibinfo {author} {\bibfnamefont {H.}~\bibnamefont {{De
  Keersmaecker}}}, \bibinfo {author} {\bibfnamefont {S.~F.}\ \bibnamefont
  {Konrad}}, \bibinfo {author} {\bibfnamefont {N.}~\bibnamefont {Kodera}},
  \bibinfo {author} {\bibfnamefont {T.}~\bibnamefont {Ando}}, \bibinfo {author}
  {\bibfnamefont {J.}~\bibnamefont {Lipfert}}, \bibinfo {author} {\bibfnamefont
  {S.}~\bibnamefont {{De Feyter}}}, \ and\ \bibinfo {author} {\bibfnamefont
  {W.}~\bibnamefont {Vanderlinden}},\ }\href@noop {} {\bibfield  {journal}
  {\bibinfo  {journal} {ACS Nano}\ }\textbf {\bibinfo {volume} {12}},\ \bibinfo
  {pages} {11907} (\bibinfo {year} {2018})}\BibitemShut {NoStop}%
\bibitem [{\citenamefont {Sankararaman}\ and\ \citenamefont
  {Marko}(2005)}]{Sankararaman2005}%
  \BibitemOpen
  \bibfield  {author} {\bibinfo {author} {\bibfnamefont {S.}~\bibnamefont
  {Sankararaman}}\ and\ \bibinfo {author} {\bibfnamefont {J.~F.}\ \bibnamefont
  {Marko}},\ }\href {\doibase 10.1103/PhysRevE.71.021911} {\bibfield  {journal}
  {\bibinfo  {journal} {Physical Review E - Statistical, Nonlinear, and Soft
  Matter Physics}\ }\textbf {\bibinfo {volume} {71}},\ \bibinfo {pages} {1}
  (\bibinfo {year} {2005})}\BibitemShut {NoStop}%
\bibitem [{\citenamefont {Brahmachari}\ \emph {et~al.}(2018)\citenamefont
  {Brahmachari}, \citenamefont {Dittmore}, \citenamefont {Takagi},
  \citenamefont {Neuman},\ and\ \citenamefont {Marko}}]{Brahmachari2018}%
  \BibitemOpen
  \bibfield  {author} {\bibinfo {author} {\bibfnamefont {S.}~\bibnamefont
  {Brahmachari}}, \bibinfo {author} {\bibfnamefont {A.}~\bibnamefont
  {Dittmore}}, \bibinfo {author} {\bibfnamefont {Y.}~\bibnamefont {Takagi}},
  \bibinfo {author} {\bibfnamefont {K.~C.}\ \bibnamefont {Neuman}}, \ and\
  \bibinfo {author} {\bibfnamefont {J.~F.}\ \bibnamefont {Marko}},\ }\href
  {\doibase 10.1103/PhysRevE.97.022416} {\bibfield  {journal} {\bibinfo
  {journal} {Phys. Rev. E}\ }\textbf {\bibinfo {volume} {97}},\ \bibinfo
  {pages} {1} (\bibinfo {year} {2018})}\BibitemShut {NoStop}%
\bibitem [{\citenamefont {Kr{\"{a}}mer}\ \emph {et~al.}(1987)\citenamefont
  {Kr{\"{a}}mer}, \citenamefont {Niem{\"{o}}ller}, \citenamefont {Amouyal},
  \citenamefont {Revet}, \citenamefont {von Wilcken-Bergmann},\ and\
  \citenamefont {M{\"{u}}ller-Hill}}]{Kramer1987}%
  \BibitemOpen
  \bibfield  {author} {\bibinfo {author} {\bibfnamefont {H.}~\bibnamefont
  {Kr{\"{a}}mer}}, \bibinfo {author} {\bibfnamefont {M.}~\bibnamefont
  {Niem{\"{o}}ller}}, \bibinfo {author} {\bibfnamefont {M.}~\bibnamefont
  {Amouyal}}, \bibinfo {author} {\bibfnamefont {B.}~\bibnamefont {Revet}},
  \bibinfo {author} {\bibfnamefont {B.}~\bibnamefont {von Wilcken-Bergmann}}, \
  and\ \bibinfo {author} {\bibfnamefont {B.}~\bibnamefont
  {M{\"{u}}ller-Hill}},\ }\href@noop {} {\bibfield  {journal} {\bibinfo
  {journal} {The EMBO journal}\ }\textbf {\bibinfo {volume} {6}},\ \bibinfo
  {pages} {1481} (\bibinfo {year} {1987})}\BibitemShut {NoStop}%
\bibitem [{\citenamefont {Fosado}\ \emph {et~al.}(2023)\citenamefont {Fosado},
  \citenamefont {Howard}, \citenamefont {Weir}, \citenamefont {Noy},
  \citenamefont {Leake},\ and\ \citenamefont {Michieletto}}]{Fosado2023}%
  \BibitemOpen
  \bibfield  {author} {\bibinfo {author} {\bibfnamefont {Y.~A.}\ \bibnamefont
  {Fosado}}, \bibinfo {author} {\bibfnamefont {J.}~\bibnamefont {Howard}},
  \bibinfo {author} {\bibfnamefont {S.}~\bibnamefont {Weir}}, \bibinfo {author}
  {\bibfnamefont {A.}~\bibnamefont {Noy}}, \bibinfo {author} {\bibfnamefont
  {M.~C.}\ \bibnamefont {Leake}}, \ and\ \bibinfo {author} {\bibfnamefont
  {D.}~\bibnamefont {Michieletto}},\ }\href@noop {} {\bibfield  {journal}
  {\bibinfo  {journal} {Physical Review Letters}\ }\textbf {\bibinfo {volume}
  {130}},\ \bibinfo {pages} {58203} (\bibinfo {year} {2023})}\BibitemShut
  {NoStop}%
\bibitem [{\citenamefont {Dittmore}\ \emph {et~al.}(2017)\citenamefont
  {Dittmore}, \citenamefont {Brahmachari}, \citenamefont {Takagi},
  \citenamefont {Marko},\ and\ \citenamefont {Neuman}}]{Dittmore2017}%
  \BibitemOpen
  \bibfield  {author} {\bibinfo {author} {\bibfnamefont {A.}~\bibnamefont
  {Dittmore}}, \bibinfo {author} {\bibfnamefont {S.}~\bibnamefont
  {Brahmachari}}, \bibinfo {author} {\bibfnamefont {Y.}~\bibnamefont {Takagi}},
  \bibinfo {author} {\bibfnamefont {J.~F.}\ \bibnamefont {Marko}}, \ and\
  \bibinfo {author} {\bibfnamefont {K.~C.}\ \bibnamefont {Neuman}},\ }\href
  {\doibase 10.1103/PhysRevLett.119.147801} {\bibfield  {journal} {\bibinfo
  {journal} {Phys. Rev. Lett.}\ }\textbf {\bibinfo {volume} {119}},\ \bibinfo
  {pages} {1} (\bibinfo {year} {2017})},\ \Eprint
  {http://arxiv.org/abs/1704.07815} {arXiv:1704.07815} \BibitemShut {NoStop}%
\bibitem [{\citenamefont {Hao}\ \emph {et~al.}(2017)\citenamefont {Hao},
  \citenamefont {Shearwin},\ and\ \citenamefont {Dodd}}]{Hao2017}%
  \BibitemOpen
  \bibfield  {author} {\bibinfo {author} {\bibfnamefont {N.}~\bibnamefont
  {Hao}}, \bibinfo {author} {\bibfnamefont {K.~E.}\ \bibnamefont {Shearwin}}, \
  and\ \bibinfo {author} {\bibfnamefont {I.~B.}\ \bibnamefont {Dodd}},\
  }\href@noop {} {\bibfield  {journal} {\bibinfo  {journal} {Nature
  Communications}\ }\textbf {\bibinfo {volume} {8}} (\bibinfo {year}
  {2017})}\BibitemShut {NoStop}%
\bibitem [{\citenamefont {Rothemund}(2006)}]{Rothemund2006}%
  \BibitemOpen
  \bibfield  {author} {\bibinfo {author} {\bibfnamefont {P.~W.~K.}\
  \bibnamefont {Rothemund}},\ }\href {\doibase 10.1038/nature04586} {\bibfield
  {journal} {\bibinfo  {journal} {Nature}\ }\textbf {\bibinfo {volume} {440}},\
  \bibinfo {pages} {297} (\bibinfo {year} {2006})}\BibitemShut {NoStop}%
\end{thebibliography}%

\end{document}